\documentclass[a4paper,12pt]{article}
\usepackage{feynmp-auto,expdlist}
\usepackage{amsmath, amsfonts}
\usepackage{graphicx,arydshln}
\usepackage{enumerate}
\usepackage{hyperref}
\usepackage{latexsym}
\usepackage{dsfont}
\usepackage{hepnicenames}
\usepackage{enumerate}
\usepackage{soul}
\usepackage[normalem]{ulem}
\usepackage{comment}

\newcommand{\footnoten}[1]{}

\usepackage[font={small},textfont={it}]{caption} 

\usepackage{units}
\newcommand{\AS}[1]{{\color{red}{[\bf AS: #1]}}}

\usepackage{mathrsfs,graphicx,rotating,amsmath,amsfonts,mathtools,booktabs,amssymb,wasysym}
\usepackage{hyperref}\usepackage{slashed}
\usepackage[nosort]{cite}
\usepackage[table,xcdraw,dvipsnames]{xcolor}
\usepackage{bm}
\usepackage{multirow,multicol}
\hypersetup{colorlinks,bookmarksopen,bookmarksnumbered,
	linkcolor=blus,pdfstartview=FitH,urlcolor=rossos,citecolor=verde}
\allowdisplaybreaks

\newcommand{\myfootnote}[1]{}
\newcommand{\myomit}[1]{{\color{gray}#1}}
\renewcommand{\myomit}[1]{}

\renewcommand{\[}{\left[}

\def\Lag{\mathscr{L}}

\newcommand{\mio}[1]{}

\def\bpm{\begin{pmatrix}}
	\def\epm{\end{pmatrix}}

\usepackage{mathrsfs}

\newcommand{\fig}[1]{~\ref{fig:#1}}

\renewcommand{\Re}{{\rm Re}\,}
\renewcommand{\Im}{{\rm Im}\,}
\allowdisplaybreaks
\usepackage{multicol}
\usepackage{color}
\definecolor{rosso}{cmyk}{0,1,1,0.4}
\definecolor{rossos}{cmyk}{0,1,1,0.55}
\definecolor{rossoc}{cmyk}{0,1,1,0.2}
\definecolor{blu}{cmyk}{1,1,0,0.3}
\definecolor{blus}{cmyk}{1,1,0,0.6}
\definecolor{bluc}{cmyk}{1,1,0,0.1}
\definecolor{verde}{cmyk}{0.92,0,0.59,0.25}
\definecolor{verdec}{cmyk}{0.92,0,0.59,0.15}
\definecolor{verdes}{cmyk}{0.92,0,0.59,0.4}

\newcommand{\bp}{\bar{M}_{\rm Pl}}
\newcommand{\eq}[1]{~{\rm (\ref{eq:#1})}}

\newcommand{\GeV}{\,{\rm GeV}}

\newcommand{\Tr}{\,{\rm Tr}}

\def\circa#1{\,\raise.3ex\hbox{$#1$\kern-.75em\lower1ex\hbox{$\sim$}}\,}

\newcommand{\nn}{\nonumber}
\newcommand{\beq}{\begin{equation}}
\newcommand{\eeq}{\end{equation}}

\newcommand{\bea}{\begin{eqnarray}}
\newcommand{\eea}{\end{eqnarray}}
\newcommand{\be}{\begin{equation}}
\newcommand{\ee}{\end{equation}}
\font\tenrsfs=rsfs10 at 12pt
\font\sevenrsfs=rsfs7
\font\fiversfs=rsfs5
\newfam\rsfsfam
\textfont\rsfsfam=\tenrsfs
\scriptfont\rsfsfam=\sevenrsfs
\scriptscriptfont\rsfsfam=\fiversfs

\newcommand{\tr}{\mathrm{Tr}}

\newsavebox\MBox

\renewenvironment{thebibliography}[1]
{\begin{multicols}{2}[\section*{\refname}]%
		\@mkboth{\MakeUppercase\refname}{\MakeUppercase\refname}%
		\list{\@biblabel{\@arabic\c@enumiv}}%
		{\settowidth\labelwidth{\@biblabel{#1}}%
			\leftmargin\labelwidth
			\advance\leftmargin\labelsep
			\@openbib@code
			\usecounter{enumiv}%
			\let\p@enumiv\@empty
			\renewcommand\theenumiv{\@arabic\c@enumiv}}%
		\sloppy
		\clubpenalty4000
		\@clubpenalty \clubpenalty
		\widowpenalty4000%
		\sfcode`\.\@m}
	{\def\@noitemerr
		{\@latex@warning{Empty `thebibliography' environment}}%
		\endlist\end{multicols}}

\newcommand{\SU}{\,{\rm SU}}

\renewcommand{\L}\Lag

\def\circa#1{\,\raise.3ex\hbox{$#1$\kern-.75em\lower1ex\hbox{$\sim$}}\,}
\makeatletter

\font\ital=cmu10

\def\hhref#1{\href{http://arxiv.org/abs/#1}{arXiv:#1}}
\usepackage{xstring}
\newcommand{\hhrefq}[1]{\IfSubStr{#1}{:}{\href{http://inspirehep.net/search?ln=en&ln=en&p=#1&of=hb&action_search=Search&sf=&so=d&rm=&rg=25&sc=0}{InSpire:#1}}{\hhref{#1}}}

\def\art{\@ifnextchar[{\eart}{\oart}}
\def\eart[#1]#2#3#4#5#6{{\rm #2}, {\em #3 \bf #4} {\rm (#6) #5} ({\em #1})}
\def\article{\@ifnextchar[{\earticle}{\oarticle}}
\def\oarticle#1#2#3#4#5#6{{\rm #1}, {\ital `#6'}, {\rm #2 #3 (#5) #4}}
\def\earticle[#1]#2#3#4#5#6#7{{\rm #2}, {\ital `#7'}, {\rm #3 #4 (#6) #5}  [\hhrefq{#1}]}
\def\hepart[#1]#2{{\rm #2, \sl#1}}
\def\heparticle[#1]#2#3{#2, {\ital `#3'} [\hhrefq{#1}]}
\newcommand{\doi}[1]{\href{http://dx.doi.org/#1}{[link]}}

\newcommand{\hhrefqq}[1]{\IfBeginWith{#1}{10.}{\href{https://doi.org/#1}{doi:#1}}{\hhrefq{#1}}}
\def\earticle[#1]#2#3#4#5#6#7{{\rm #2}, {\ital `#7'}, {\rm #3 #4 (#6) #5}  [\hhrefqq{#1}]}

\renewenvironment{thebibliography}[1]
{\begin{multicols}{2}[\section*{\refname}]%
		\@mkboth{\MakeUppercase\refname}{\MakeUppercase\refname}%
		\list{\@biblabel{\@arabic\c@enumiv}}%
		{\settowidth\labelwidth{\@biblabel{#1}}%
			\leftmargin\labelwidth
			\advance\leftmargin\labelsep
			\@openbib@code
			\usecounter{enumiv}%
			\let\p@enumiv\@empty
			\renewcommand\theenumiv{\@arabic\c@enumiv}}%
		\sloppy
		\clubpenalty4000
		\@clubpenalty \clubpenalty
		\widowpenalty4000%
		\sfcode`\.\@m}
	{\def\@noitemerr
		{\@latex@warning{Empty `thebibliography' environment}}%
		\endlist\end{multicols}}

\newcounter{alphaequation}[equation]
\def\thealphaequation{\theequation\hbox to
	0.6em{\hfil\alph{alphaequation}\hfil}}
\def\eqnsystem#1{
	\def\@eqnnum{{\rm (\thealphaequation)}}
	\def\@@eqncr{\let\@tempa\relax \ifcase\@eqcnt \def\@tempa{& & &} \or
		\def\@tempa{& &}\or \def\@tempa{&}\fi\@tempa
		\if@eqnsw\@eqnnum\refstepcounter{alphaequation}\fi
		\global\@eqnswtrue\global\@eqcnt=0\cr}
	\refstepcounter{equation} \let\@currentlabel\theequation \def\@tempb{#1}
	\ifx\@tempb\empty\else\label{#1}\fi
	\refstepcounter{alphaequation}
	\let\@currentlabel\thealphaequation
	\global\@eqnswtrue\global\@eqcnt=0 \tabskip\@centering\let\\=\@eqncr
	$$\halign to \displaywidth\bgroup \@eqnsel\hskip\@centering
	$\displaystyle\tabskip\z@{##}$&\global\@eqcnt\@ne
	\hskip2\arraycolsep\hfil${##}$\hfil& \global\@eqcnt\tw@\hskip2\arraycolsep
	$\displaystyle\tabskip\z@{##}$\hfil
	\tabskip\@centering&\llap{##}\tabskip\z@\cr}
\def\endeqnsystem{\@@eqncr\egroup$$\global\@ignoretrue} \makeatother

\definecolor{Gray}{gray}{0.95}

\def\bal#1\eal{\begin{align}#1\end{align}}

\begin{document}
\thispagestyle{empty}
\vspace{0.1cm}
\begin{center}
{\LARGE \bf \color{rossos} Modular invariance and the QCD angle}\\[2ex]
\vspace{1cm}
{\bf\large Ferruccio Feruglio$^a$, Alessandro Strumia$^b$, Arsenii Titov$^b$}  \\[5mm]
{$^a$ \em INFN, Sezione di Padova, Italia}\\
{$^b$ \em Dipartimento di Fisica, Universit{\`a} di Pisa, Italia}\\[4ex]

\begin{center}{\bf\color{blus} Abstract}
\begin{quote}
\large
String compactifications on an orbi-folded torus 
with complex structure
give rise to chiral fermions, spontaneously broken CP, modular invariance.
We show that this allows simple effective theories of flavour and CP where: 
i) the QCD angle vanishes;
ii) the CKM phase is large; 
iii) quark and lepton masses and mixings can be reproduced up to order one coefficients.
We implement such general paradigm in supersymmetry or supergravity, with modular forms or functions,
with or without heavy colored states.
\end{quote}
\end{center}
\end{center}

\setcounter{tocdepth}{1}
\tableofcontents

\normalsize

\section{Introduction}
Data show that CP is violated by the order-unity  phase $\delta_{\rm CKM} \sim 1$ in the CKM matrix,
while the upper bound on the neutron electric dipole~\cite{2001.11966} implies the smallness of the QCD angle
\beq\bar\theta= \theta_{\rm QCD} + \arg\det M_q,\qquad | \bar\theta|\circa{<} 10^{-10}\eeq
where $M_q$ is the mass matrix of quarks $q$ and
$\theta_{\rm QCD}$ is the coefficient of the topological term in the QCD Lagrangian
\beq
\Lag_{\rm QCD} = \bar q (i \slashed{D} - M_q)q - \frac14 \Tr \, G^2  + \theta_{\rm QCD} \frac{g_3^2}{32\pi^2} \Tr\, G\tilde{G}.\eeq
This aspect of the Standard Model is puzzling, 
because a generic complex $M_q$  leads to $\delta_{\rm CKM},\bar\theta \sim 1$. 
This puzzle has been interpreted in two different ways:
\begin{enumerate}[a)]
\item as a new light pseudo-scalar, the axion~\cite{PQ,Luzio}, adjusting $\bar\theta=0$ dynamically;
the axion is so far not observed;
special models are needed to avoid the `axion quality problem'~\cite{Luzio};

\item as special models that produce a large $\delta_{\rm CKM}$, 
a quark mass matrix $M_q$ with real determinant and $\theta_{\rm QCD} =0$.
\end{enumerate}
Models of type b) have been first realized by Nelson and Barr~\cite{Nelson:1983zb,Barr:1984qx} by the ad-hoc assumption 
that CP is violated only by the mixings 
of SM quarks $q$ with hypothetical extra heavy quarks, and that the extended
quark mass matrix has a special structure with vanishing entries
such that CP-violating terms don't contribute to its determinant.
To avoid that higher-order corrections violate the needed structure, 
these models seem to need to operate at relatively low scale and
supersymmetry broken at low energy in a CP-conserving way, such as gauge mediation~\cite{1506.05433}.
Other models assume real Yukawas and a complex kinetic matrix in supersymmetry~\cite{hep-ph/0105254},
parity suitably broken~\cite{Babu:1989rb,Kuchimanchi:1995rp},
mirror sectors that duplicate the SM~\cite{Barr:1991qx,2303.06156};
texture zeroes enforced via complicated patterns of symmetry breaking~\cite{1307.0710},
warped extra dimensions with extra structure~\cite{hep-ph/0411132,0711.4421,1412.3805}.
Problems with Planck-suppressed operators can be avoided assuming that these mechanisms operate at low enough energy~\cite{hep-ph/9212318,2106.09108}.

\medskip

QFT models where CP is imposed and broken cannot however go to the heart of the problem: 
exploring if a theory that provides a fundamental origin of CP can select the special configuration $\bar \theta \ll \delta_{\rm CKM}$.
A candidate is string theory, where chiral fermions, CP and its violation can arise geometrically 
from compactifications on a 6-dimensional
space with complex structure (e.g.~\cite{hep-th/9205011,hep-ph/9205202}).
In the effective QFT below the string scale this physics is described by CP-violating scalar moduli that control
the shape of the compactification space.
In $N=1$ supersymmetric toroidal compactifications the moduli that remain after orbi-folding
enjoy a special {\em modular invariance} of stringy origin~\cite{Dixon:1985jw,Dixon:1986jc}.
Modular invariance is special because it arises as a symmetry of how strings experience the geometry of the compactification space,
and more generically because an infinite number of heavy states are integrated out. 
It strongly constrains 
interactions among states, both at the string level~\cite{Hamidi:1986vh,Dixon:1986qv,Lauer:1989ax,Lauer:1990tm} and in the low-energy regime~\cite{Ferrara:1989bc,Ferrara:1989qb}.

\smallskip

Modular invariance has been recently studied, independently of its string motivation,
to build more predictive flavour models for neutrino, lepton and quark masses~\cite{1706.08749}.
This profits from the fact that finite copies of the modular group are isomorphic to
the non-Abelian finite symmetries previously
used in neutrino flavour model-building.
Such symmetries are broken in a specific way that can be predictive, since the complicated symmetry-breaking sector of the earlier model-building reduces to a single complex field, the modulus.
For the purpose of our present discussion, 
we do not need to consider finite modular symmetries.

\smallskip

In section~\ref{QCDmod} we discuss how one modular symmetry non-anomalous under QCD neatly 
explains $\bar\theta\ll \delta_{\rm CKM}$ in the minimal MSSM with global supersymmetry,
even allowing for non-minimal kinetic terms.
In section~\ref{FN} we `deconstruct' modular models for $\bar\theta\ll \delta_{\rm CKM}$, 
showing how the key ingredients automatically provided by modular invariance 
could be artificially implemented in simpler U(1) Froggatt-Nielsen-like models~\cite{Froggatt:1978nt}.
In section~\ref{poles} we implement the same basic idea in MSSM extensions with optional heavy quarks: 
integrating them out leads to more general effective theories with apparently anomalous modular symmetry
and with modular forms replaced by singular modular functions, that still explain $\bar\theta\ll \delta_{\rm CKM}$.
Section~\ref{sugra} shows that the mechanism can be extended to supergravity,
where the gluino gets involved in modular transformations, making heavy colored states needed. 
In section~\ref{string} we discuss the possibility of identifying such states as string states,
speculating that the proposed mechanism for $\bar\theta\ll \delta_{\rm CKM}$ could arise in string theory.
In section~\ref{Corrections} we discuss supersymmetry breaking and other effects that shift $\bar\theta$ away from 0.
Conclusions are given in section~\ref{concl}.

\section{Modular invariance and global supersymmetry}\label{QCDmod}

\subsection{Modular invariance}\label{modular}
Here we consider an extension of the Standard Model with $N=1$ global supersymmetry.
As usual the SM quarks are part of chiral multiplets $\Phi = \{Q,u_R,d_R\}$, 
and two Higgs doublets appear in chiral multiplets $\Phi=\{H_u, H_d\}$.\footnote{In our notation they all contain
left-handed Weyl fermions. In particular our $q_R$ corresponds to what is more commonly denoted as
$q_R^c$ or as $q^c$.}
We assume an extra complex modulus $\tau$, the scalar component of another chiral multiplet $\tau$.
This is motivated by string theory, where $\tau$ controls the geometry of the complex  compactification space.
For simplicity we assume a single $\tau$;  toroidal super-string compactifications tend to give multiple modular symmetries and moduli.
We ask the full  low-energy effective physics to be invariant under 
\beq\label{eq:modtau} \tau\to \frac{a\tau+b }{c\tau+d}\eeq
where $a,b,c,d$ are integers with $ad-bc=1$.
This defines the action of the modular group SL(2,~$\mathbb{Z}$) on the modulus $\tau$.
We assume that the vacuum expectation value of $\tau$ is fixed by some mechanism~\cite{Cvetic:1991qm,1812.06520,2201.02020,2304.14437},
though we do not need a special value of $\tau$.
Whatever value $\tau$ takes, modular invariance is spontaneously broken by $\tau$ in a predictive way.
In technical language, the modular symmetry is non-linearly realized.
The modulus $\tau$ takes values in the upper half of the complex plane. In a modular-invariant theory,
we can further restrict this region to the fundamental domain, see fig.\fig{E2E4plot}.

\medskip

The  K\"ahler potential $K(\Phi,\Phi^\dagger)$ (describing the kinetic terms), the super-potential $W(\Phi)$ (describing the Yukawa couplings) and the gauge kinetic function $f$ have the following minimal form:
\begin{eqnsystem}{sys:KW}  
\label{eq:K}
K &=&  -h^2 \ln(-i\tau+i\tau^\dagger) + \sum_{\Phi} \frac{\Phi^\dagger {e^{2 V}} \Phi}{(-i\tau + i \tau^\dagger)^{k_\Phi}},\\
W&=&Y^u_{ij}(\tau) H_{u} u_{Ri} Q_{j} +Y^d_{ij}(\tau) H_{d}  d_{Ri} Q_j,\\
f&=&f_0,
\end{eqnsystem}
where the `weight' $k_\Phi$ of $\Phi$ is a number (a rational number in string theory~\cite{hep-th/9202046,Ferrara:1989qb})
and $f_0$ is a constant.
Similarly to the axion decay constant, $h$ sets the scale of the dimension-less field $\tau$.
The effective theory must be invariant under the modular group SL$(2,\mathbb{Z})$.
The kinetic term of $\Phi$ is invariant if the vector multiplet $V$ is invariant and $\Phi$ transforms as 
\beq\label{eq:modPhi}
\Phi \to (c\tau+d)^{-k_\Phi} {\rho_\Phi\cdot \Phi}
\eeq
with $\rho_\Phi$ a phase, possibly depending on $a$, $b$, $c$, $d$.\footnote{A non-trivial matrix $\rho_\Phi$ 
that represents 
a finite modular symmetry appears when considering a multiplet $\Phi$ and
normal sub-groups of SL$(2,\mathbb{Z})$ of level higher than 1.
Quotients of SL$(2,\mathbb{Z})$ with respect to these normal subgroups are finite non-Abelian groups.
For simplicity we here focus on the full modular group with level 1, where $\rho_\Phi$ is just a phase.}
This applies to SM quarks and Higgses $H_{u,d}$. 
The first term in $K$ describes the special kinetic term for $\tau$ and transforms as
\beq \ln(-i\tau+i \tau^\dagger) \to  \ln(-i\tau+i \tau^\dagger) - \ln(c\tau+d)-\ln(c\tau^\dagger + d)
\label{eq:Ktransform}\eeq
leaving $K$ invariant up to a K\"ahler
transformation $K(\tau,\tau^\dagger)\to K(\tau,\tau^\dagger)+ g(\tau)+g(\tau^\dagger)$ that has no effect in  global supersymmetry.
The K\"ahler potential of eq.~(\ref{eq:K}) is not the most general one allowed by modular invariance and
can be easily generalized to include non-minimal (and non-diagonal) terms. Although in our discussion we 
adopt the minimal form of eq.~(\ref{eq:K}), our results are not modified by choosing the most general
K\"ahler potential, as shown in section \ref{CP}.

\smallskip

The Yukawa couplings $Y$ must depend on the super-field $\tau$ such that the effective theory is modular invariant.
This means that each entry $Y(\tau)$ of the Yukawa coupling matrices $ Y^q_{ij}(\tau)$ must transform as\footnote{If non-trivial phases $\rho_\Phi$ are present, they should sum up to zero in each term of $W$, see appendix~\ref{phases}.}
\beq Y(\tau)\to (c\tau+d)^{k_Y} Y(\tau)\eeq
with weight $k_Y = k_{q_{Ri}}+k_{q_{Lj}} + k_{H_q} - k_W$. 
Here $k_W=0$ is the modular weight of
the super-potential $W$ (it will be non-vanishing in supergravity),
and $k_{H_{u,d}}$ are the modular weights of the Higgs doublets $H_{u,d}$.
The functional dependence of the Yukawa couplings on $\tau$ must be
\beq \label{eq:Yij}
Y^q_{ij}(\tau) = c^q_{ij} \, F_{k_{ij}}(\tau),\qquad k_{ij} =k_{q_{Ri}} + k_{q_{Lj}}+k_{H_q} \eeq
where the functions $F_k(\tau)$ with fixed modular weight $k$ are nearly unique if
we assume they are holomorphic everywhere in the fundamental region of SL$(2,\mathbb{Z})$,
including the point at infinity $\tau=i\infty$. This assumption will be critically discussed in section~\ref{poles}.
Functions with these properties are modular forms and they only exist if $k$ is a non-negative even integer:
$F_0=1$ is a constant, $F_2$ vanishes, the first non-trivial forms are $F_4 = E_4$ and $F_6=E_6$.
Here $E_{k}(\tau)$ are the holomorphic normalized Eisenstein series with given modular weight $k$ thanks to the 
lattice summation over pairs of integers $m,n$:
\beq E_{k}(\tau) = \frac{1}{2\zeta(k)} \sum_{(m,n)\neq(0,0)}\frac{1}{(m + n\tau)^{k}}\eeq
where the normalization factor makes $E_{4,6,\ldots}\simeq 1$ at large $\Im\tau$.
Some known mathematical results: $E_2$ is divergent and cannot be cured.
The series $E_{4,6}$ generate the whole set of modular forms: a generic form $F_k$ is a polynomial in
$E_{4,6}$. Therefore, going to higher orders one has $E_8=E_4^2$ and $E_{10}=E_4 E_6$, see table~\ref{tab:basis}.
The forms $E_4(\tau)$ and $E_6(\tau)$ are plotted in fig.\fig{E2E4plot} and one crucial property is that 
$E_4^3\neq E_6^2$ have different phases. 
The most generic function with weight 12 is a linear combination $E_4^3 + c E_6^2$ where $c$ is a constant.
The Eisenstein form $E_{12}$ corresponds to a specific value of $c$, not needed for our purposes.
In general $N+1 $ multiple functions appear at weight $12 N$ where $N$ is a positive integer: 
e.g.\ at weight 24 one has $E_4^6 ,E_6^4 $ and $ E_4^3 E_6^2$.

\begin{table}[t]
\centering
\begin{tabular}{|l|cccccccc|}
\hline
Modular weight $k$ & 0 & 2 & 4 & 6 & 8 & 10 & 12 & 14 \\
\hline
Number of forms & 1 & 0 & 1 & 1 & 1 & 1 & 2 & 1 \\
\hline
Modular forms & 1 & -- & $E_4$ & $E_6$ & $E_8 = E_4^2$ & $E_{10} = E_4 E_6$ & $E_{4}^3, E_6^2$ & $E_{14}=E_4^2 E_6$ \\
\hline
\end{tabular}
\caption{\label{tab:basis} Modular forms up to weight $k \leq 14$.
}

\end{table}

\subsection{CP invariance}\label{CPI}
\begin{figure}[t]
$$\includegraphics[width=\textwidth]{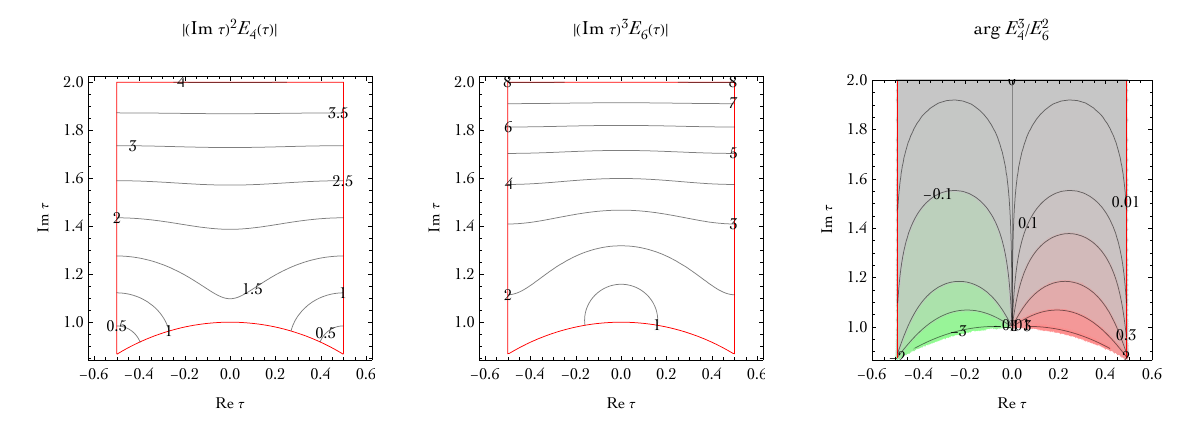}$$
\caption{\label{fig:E2E4plot} Absolute values of the modular forms $E_4$ (left plot) and $E_6$ (middle) and their relative phase in the physical
domain for the modulus $\tau$ (right). }
\end{figure}

We choose a basis in field space where CP transformations act on $\tau$ and $\Phi$ as
\beq
\tau\to- \tau^\dagger \qquad \Phi\to \Phi^\dagger.
\eeq
The Eisenstein functions satisfy $E_{k}(-\tau^\dagger)=E_{k}(\tau)^\dagger$.
We focus on a CP-invariant theory~\cite{1905.11970,Baur:2019kwi}, where $c^{u,d}_{ij}$ in eq.~(\ref{eq:Yij}) are real, $F_k$ are polynomials in $E_{4,6}$ with real coefficients and $\theta_{\rm QCD}=0$.
Then CP invariance can only be broken spontaneously 
and we assume that the only source of CP violation is the vacuum expectation value of $\tau$.
This occurs for a generic value of $\tau$ not lying along the imaginary $\tau$ axis, nor along
the border of the fundamental region of fig.\fig{E2E4plot}.

\subsection{CP violation: solving the $\bar\theta\ll\delta_{\rm CKM}$ puzzle}\label{CP}

We now show how supersymmetric CP and modular-invariant theories can
easily produce Yukawa couplings such that the CKM phase is large and the QCD ${\bar\theta}$ angle vanishes.
\smallskip
The relative phase between $E_4^3$ and $E_6^2$ allows to induce a physical CP-violating phase in the Yukawa matrices,
giving the CKM phase and possibly a contribution to $\bar\theta$ from the phase of the quark masses,
$\arg \det M_q $. 
In such a case $\Re\tau$ would be an axion that dynamically adjusts $\bar\theta=0$ if the $\tau$ potential were dominated by QCD effects.
However, supersymmetry-breaking effects are expected to dominate~\cite{Ibanez:1991qh},
as no global continuous $U(1)_{\rm PQ}$ symmetry keeps $\tau$ light.
Theories with modular invariance and an axion were considered in~\cite{2002.06931}.

\smallskip

A different case that does not need this assumption is possible.
The key observation is that
modular transformations are multiplicative,  so $\det Y_q$ is a modular form with modular weight 
$\sum_{i=1}^3 (k_{q_{Li}} +k_{q_{Ri}}+k_{H_q})$. 
The Higgs bosons acquire vacuum expectation values that would break the modular symmetry:
to avoid this we assume $k_{H_u}=k_{H_d}=0$ 
(the weaker condition $k_{H_u}+k_{H_d}=0$ would be enough for our purposes),
so that $\arg \det M_q = \arg \det Y_u\det Y_d$.
Our final assumption needed to solve the QCD $\theta$ problem is
\beq \label{eq:k0} A \equiv \sum_{i=1}^3 (2 k_{Q_i}+ k_{u_{Ri}}+k_{d_{Ri}}) =0.\eeq
This guarantees that:
\begin{enumerate}
\item the modular symmetry has no QCD anomaly $A$:
a $\tau$-dependent redefinition of the phases of quark super-fields does not affect the kinetic function of the gluon super-multiplet;
\item the product $\det M_q = \det M_u M_d$ of all quark masses is real,
as it is a $\tau$-independent modular form with weight 0.
\end{enumerate}
We thereby have $\theta_{\rm QCD}=0$ and real $\det M_q$.
The real constants $c^q_{ij}$ can be chosen such that $\det M_q>0$, yielding $\bar\theta=0$ rather than $\bar\theta=\pi$.
Finally, $\bar\theta=0$ is preserved provided that the gluino mass is real: this is satisfied by any mechanism 
of supersymmetry breaking that preserves CP.
This can be easily realized by assuming that supersymmetry is broken in a sector with vanishing modular charges.

\smallskip

The condition of eq.\eq{k0} allows for CP-violating Yukawa matrices, as long as
they depend on both $E_4$ and $E_6$.
Generally speaking, the
eigenvalues and eigenvectors of the quark mass matrices depend non trivially on all quark weights.
Specifically, a non-vanishing CKM phase is signalled by the Jarlskog invariant
$\Im\det [Y_u^\dagger Y_u, Y_d^\dagger Y_d]\neq 0$~\cite{Jarlskog:1985ht} for $3$ quark generations, 
and more generally by $\Im\Tr [Y_u^\dagger Y_u, Y_d^\dagger Y_d]^3 \neq 0$~\cite{Bernabeu:1986fc}.
These combinations are invariant under quark field redefinitions, but are not holomorphic and have no special modular properties.

\subsubsection*{Wave function renormalization}
Eq.\eq{Yij} for the Yukawa couplings holds in a basis where the minimal kinetic term of quarks in eq.\eq{K} is non-canonical.
The canonically normalized superfield $\Phi_{\rm can}=\Phi/(-i\tau + i\tau^\dagger)^{k_\Phi/2}$ transforms 
acquiring a phase $\Phi_{\rm can}\to [(c\tau+d)/(c\tau^\dagger+d)]^{-k_\Phi/2} \Phi_{\rm can}$ under a modular transformation.
The Yukawa matrices for canonically normalized quarks are 
\beq Y^q_{ij}|_{\rm can} = c_{ij}^q (2\Im\tau)^{k_{ij}/2} F_{k_{ij}}(\tau)\qquad\hbox{where, again}\qquad 
k_{ij} =k_{q_{Ri}} + k_{q_{Lj}} +k_{H_q}.\eeq
Furthermore, extra non-minimal kinetic terms are possible, because
the $3\times 3$ kinetic matrices $Z_f(\tau,\tau^\dagger)$ of fermions $f =\{u_R, d_R, Q\}$ are not holomorphic in $\tau$,
and modular invariance allows them to depend on the CP-violating parameters $\tau,\tau^\dagger$ in new ways. 
These non-minimal kinetic terms reduce the predictive power of flavour models based on modular symmetries~\cite{1706.08749,1909.06910,2101.08718,2108.02240}
and are often assumed to be negligible.

Such extra complex terms are not a problem for our proposed interpretation of the QCD problem, $\bar \theta=0$. 
Indeed each kinetic matrix $Z_f$  can be brought to canonical form via a general linear transformation of the three generations of
$f_{1,2,3}$ quarks:
a linear transformation affects both $\arg\det M_q$ and $\theta_{\rm QCD}$  (via the anomaly) but leaves the physical combination $\bar\theta$ invariant.
Furthermore, these linear transformations can be chosen in ways that 
leave $\arg\det M_q$ and $\theta_{\rm QCD}$ separately invariant,
by decomposing each kinetic matrix $Z_f$ either as $Z_f=H_f^\dagger H_f$ (where
$H_f$ is an hermitian matrix, see e.g.~\cite{Dugan:1984qf}) or as 
$Z_f = V_f^\dagger  \Delta_f^2 V_f$ (where $\Delta_f$ is a diagonal matrix with real positive entries
and $V_f$ is a product of 3 complex rotations with unit determinant).
The consequent linear transformation of quark fields affects their masses and mixings (including the CKM phase)
without affecting $\arg\det M_q$.

This discussion shows that, unlike fermion masses and mixing angles, the physical $\bar\theta$ angle is a holomorphic quantity completely
insensitive to the K\"ahler potential and can be effectively constrained
by modular invariance alone, at least in the limit of unbroken supersymmetry.

\begin{table}[t]
$$\begin{array}{c|ccc|ccc}
\hbox{Yukawa matrices}  & \multicolumn{3}{c}{\hbox{Modular weights}} & \multicolumn{3}{c}{\hbox{Alternative bigger weights}} \cr
Y_{u,d} &(u_L,d_L)_{1,2,3} & u_{R1,2,3} & d_{R1,2,3}&(u_L,d_L)_{1,2,3} & u_{R1,2,3} & d_{R1,2,3}\cr \hline
\begin{pmatrix} 0 & 0 & 1 \cr 0&1 & E_6 \cr  1 & E_6 & E_4^3+ E_6^2\end{pmatrix} &
\begin{pmatrix}-6 \cr 0 \cr 6\end{pmatrix}& 
\begin{pmatrix}-6 \cr 0 \cr 6\end{pmatrix}& 
\begin{pmatrix}-6 \cr 0 \cr 6\end{pmatrix}& 
\begin{pmatrix}-4 \cr 2 \cr 8\end{pmatrix}& 
\begin{pmatrix}-8 \cr -2 \cr 4\end{pmatrix}& 
\begin{pmatrix}-8 \cr -2 \cr 4\end{pmatrix}\cr
\begin{pmatrix} 0 & 0 & 1 \cr 0&1 & E_4^2 \cr  1 & E_4 & E_4^3+ E_6^2\end{pmatrix} &
\begin{pmatrix}-6 \cr -2 \cr 6\end{pmatrix}& 
\begin{pmatrix}-6 \cr 2 \cr 6\end{pmatrix}& 
\begin{pmatrix}-6 \cr 2 \cr 6\end{pmatrix}& 
\begin{pmatrix}-4 \cr -4 \cr 8\end{pmatrix}& 
\begin{pmatrix}-8 \cr 4 \cr 4\end{pmatrix}& 
\begin{pmatrix}-8 \cr 4 \cr 4\end{pmatrix}\cr \hline
\begin{pmatrix} 0 & 0 & 1 \cr 0&1 & E_4^2 \cr  1 & E_4^2& E_4(E_4^3+ E_6^2)\end{pmatrix} & &&&
\begin{pmatrix}-8 \cr 0 \cr 8\end{pmatrix}& 
\begin{pmatrix}-8 \cr 0 \cr 8\end{pmatrix}& 
\begin{pmatrix}-8 \cr 0 \cr 8\end{pmatrix}\cr  
\begin{pmatrix} 0 & 0 & 1 \cr 0&1 & E_4 E_6 \cr  1 & E_6 & E_4(E_4^3+ E_6^2)\end{pmatrix} & &&&
\begin{pmatrix}-8 \cr -2 \cr 8\end{pmatrix}& 
\begin{pmatrix}-8 \cr 2 \cr 8\end{pmatrix}& 
\begin{pmatrix}-8 \cr 2 \cr 8\end{pmatrix}\cr  
\begin{pmatrix} 0 & 0 & 1 \cr 0&1 &  E_4^3+ E_6^2  \cr  1 & E_4& E_4(E_4^3+ E_6^2)\end{pmatrix} & &&&
\begin{pmatrix}-8 \cr -4 \cr 8\end{pmatrix}& 
\begin{pmatrix}-8 \cr 4 \cr 8\end{pmatrix}& 
\begin{pmatrix}-8 \cr 4 \cr 8\end{pmatrix}\cr  
\end{array}$$
\caption{\label{tab:Y000}
Simplest modular weights that lead to Yukawa matrices such that
$\bar\theta=0$ and $\delta_{\rm CKM}\neq 0$. 
The list is complete up to permutations and transpositions, 
and assumes vanishing modular weights of the Higgs doublets and of the super-potential.
Real constants $c^q_{ij}$ are here omitted.}
\end{table}

\subsection{Concrete models for quark masses and mixings}\label{concrete}
A list of specific choices of modular weights that satisfy the above requirements is shown in table~\ref{tab:Y000}.
Solutions where all quarks are massive exist thanks to CKM mixing.
In all solutions $Y_u$ and $Y_d$ have the same non-diagonal structure, so that
$\det M_u$ and $\det M_d$ are separately real.
Sorting quark generations in increasing order of modular weights we find
\beq \label{eq:Ycan} Y_{q}|_{\rm can} = 
\bordermatrix{ & q_{L1} & q_{L2} & q_{L3}\cr 
q_{R1} & 0 & 0 & c^q_{13} \cr 
q_{R2} & 0 &c^q_{22} & c^q_{23} (2\Im\tau)^{k_{23}/2} F_{k_{23}}(\tau)\cr 
q_{R3} & c^q_{31} & c^q_{32}  (2\Im\tau)^{k_{32}/2} F_{k_{32}}(\tau) &(2\Im\tau)^{k_{33}/2} \left[c^{q}_{33} F_{k_{33}}(\tau) 
+ {c'}^{q}_{33}  F'_{k_{33}}(\tau)\right]},
\eeq
with $q=\{u,d\}$, $k_{33}=k_{23}+k_{32}$ 
and $F_k , F'_k$ denoting independent modular forms of weight $k$.
The simplest model has $k_{23}=k_{32}=6$ 
from modular weights $k_Q = k_{u_R} = k_{d_R} = (-6,0,6)$.
All modular anomalies vanish and the SM gauge group could be extended to SU(5) or SO(10) unification.\footnote{Cancellation of mixed modular anomalies with other factors of the SM gauge group is not needed for our purposes and
would require also 
$\sum_i (3k_{Q_i}+k_{L_i})=0$
and $\sum_i (k_{Q_i}+8 k_{u_{Ri}}+2 k_{d_{Ri}}+3k_{L_i}+6k_{e_{Ri}})=0$ (assuming $k_{H_u}+k_{H_d}=0$).
All anomalies cancel in the minimal model where the three generations of fermions have modular weights
$-6$, $0$ and $+6$ respectively.
}
The Yukawa couplings are given by the combination of modular forms illustrated in the upper row of table~\ref{tab:Y000}.\footnote{Similar Yukawa matrices
motivated by the QCD $\theta$ problem have been considered in~\cite{hep-ph/9612396} and in \cite{1307.0710},
where they are obtained imposing supersymmetry and a
$A_4\otimes {\rm U}(1)_R\otimes\mathbb{Z}_2\otimes\mathbb{Z}_4\otimes\mathbb{Z}_4\otimes\mathbb{Z}_4\otimes\mathbb{Z}_4\otimes\mathbb{Z}_4$ symmetry suitably broken by 9 scalar flavons.}
In this model $Y_{23}$ and $Y_{32}$ have the same modular charge.
In a different model  
$Y_{23}$ and $Y_{32}$ have different modular charges $k_{23}=8$ and $k_{32}=4$, so that  
giving the modular forms shown in the second row of table~\ref{tab:Y000}.
This model can be realized with modular charges
$k_Q = (-6,-2,6)$,  $k_{u_R} = k_{d_R} = (-6,2,6)$ as well as with different bigger values.
The less minimal models listed in table~\ref{tab:Y000} need bigger values of the modular charges.

\medskip

We next verify that such models can reproduce the observed quark masses and mixings, in addition
to $\bar\theta=0$. 
All models predict quark Yukawa matrices $Y_u$ and $Y_d$ with vanishing 11, 12 and 21 entries.
One such matrix $Y$ contains one physical phase, that can be rotated for example only into its $33$ element.
We thereby diagonalise a Yukawa matrix of the form
\beq Y =\bordermatrix{ & q_{L1} & q_{L2} & q_{L3}\cr 
q_{R1} & 0 & 0 & y_{13} \cr 
q_{R2} & 0 &y_{22} & y_{23}\cr 
q_{R3} & y_{31} & y_{32} & y_{33} e^{- i \delta}}.\eeq
The determinant is $\det Y = -y_{13}y_{31} y_{22} $.
Simple analytic expressions for masses and mixings arise in the limit where {\em all} mixing angles are small. 
The eigenvalues are
\beq y_3 \simeq y_{33},\qquad y_2\simeq y_{22},\qquad y_1\simeq -\frac{y_{13}y_{31}}{y_{33}}.\eeq
The mixing angles among left-handed quarks  are 
\beq \theta_{23}\simeq \frac{y_{32}}{y_{33}},\qquad \theta_{13}\simeq  \frac{y_{31}}{y_{33}},
\qquad \theta_{12}\simeq  \frac{y_{31} y_{23}}{y_{22}y_{33}} \eeq
and the CKM-like phase is $\delta$.
Notice that $y_{13}$ and $y_{23}$ only control $y_1$ and $\theta_{12}$, that can be computed by integrating out the
heaviest eigenvalue.
The CKM phase is observed to be large, $\delta_{\rm CKM}\approx 1.2$.
In the present theory this comes from the relative phase between 
$E_6^2$, $E_4^3$ and $E_4^3 + c E_6^2$, that thereby has to be large.
Fig.\fig{E2E4plot} shows that the phase of $E_4^3/E_6^2$ vanishes on the boundary of the fundamental domain, 
and gets small at $\Im\tau\gg 1$, where $E_4\simeq E_6\simeq 1$. 
So reproducing the CKM phase either needs $\Im\tau\sim 1$ or a value of $c \approx -1$ that gives a mild cancellation
(this cancellation could also help explaining $m_b \ll m_t$).

The Yukawa matrices can be diagonalised as
$Y_q = V_{q_R}\cdot{\rm diag}\,(y_{q1},y_{q2},y_{q3})\cdot V_{q_L}$
so that $V_{\rm CKM} = V_{u_L}\cdot V_{d_L}^\dagger$.
Assuming that the dominant contribution to the CKM matrix comes
from the mixing $V_{d_L}$ (as down quarks experimentally exhibit a milder mass
hierarchy than up quarks), the above relations can be inverted obtaining $Y_d$ in terms
of the observed down-quark Yukawas $y_{d,s,b}$ and CKM mixings,
\beq V_{\rm CKM} =
R_{23}(\theta_{23}) \cdot
\hbox{diag}\,(1,  1,e^{i \delta_{\rm CKM}})\cdot
R_{13}(\theta_{13}) \cdot
\hbox{diag}\,(1,  1,e^{-i  \delta_{\rm CKM}}) \cdot
R_{12}(\theta_{12}) .\eeq
The result is
\beq |Y_d| \simeq
\begin{pmatrix} 0 & 0 & y_{d}/\theta_{13}  \cr 
0 &y_{s} &y_{s}\theta_{12}/\theta_{13} \cr 
y_{b}\theta_{13} & y_{b}\theta_{23} & y_{b}
\end{pmatrix}\approx y_b
\begin{pmatrix} 
0 & 0 & 0.2 \cr 
0 &0.02 & 1\cr 
0.004  & 0.04& 1
\end{pmatrix},\qquad \delta_{\rm CKM} = \delta.
\eeq
In this limit, data indicate that the right-handed angle in the 23 sector might not be small.

\subsection{Numerical example}
Having understood the main result, we perform a precise numerical diagonalisation of $Y_u$ and $Y_d$
and a global fit to all quark masses and mixings.
As they can be reproduced exactly, we search for special fits such that
all constants $c_{ij}^u$ and $c_{ij}^d$ are of order unity, and the modular symmetry explains the large
$\delta_{\rm CKM}$, $\bar\theta=0$ as well as the hierarchies in quark masses and mixings.

As a numerical example, we consider the simplest model in the upper row of table~\ref{tab:Y000},
corresponding to eq.\eq{Ycan} with $k_{23}=k_{32}=6$.
The model contains 16 real  and 1 complex free parameters ($\tau,\tan\beta, c_{ij}^q$), 
that can be used to exactly reproduce
the values of the 9 real (six quark masses, three mixing angles) and 1 complex (the CKM phase) observables.
We fix $\tan\beta=10$ and $\tau = 1/8+ i$ and search for comparable values of the 
14 $c^q_{ij}$ parameters that fit values renormalized around the unification scale of $2~ 10^{16}\GeV$~\cite{1306.6879,2012.13390}.
The possible choice 
\beq {c^u_{ij}} \approx 10^{-3} \begin{pmatrix}
 0 & 0 & 1.56\\
 0 & -1.86 & 0.87 \\ 
 1.29 & 4.14 & 3.51, 1.40
 \end{pmatrix},\qquad
{ c^d_{ij} }\approx 10^{-3}  \begin{pmatrix}
 0 & 0 & 1.55\\
 0 & -2.59 & 4.59 \\ 
 0.378 & 0.710 & 0.734, 1.76
 \end{pmatrix}
 \label{eq:cqij}
\eeq
demonstrates how  modular forms can also explain the observed quark mass hierarchies in terms of order one
factors, similarly to what was achieved by Froggatt and Nielsen~\cite{Froggatt:1978nt}.
The mild expansion parameter is built in modular forms, such as the 6 in $E_6$.
For example, the factor that makes the third generation canonical is $(2\Im \tau)^6=64$ in our example.
The overall factor $10^{-3}$ was assumed in eq.\eq{cqij} because it is the typical loop factor of SM couplings;
in string compactifications the overall size of couplings is controlled by the dilaton vacuum expectation value.
The numerical example contains no special tunings.
An order unity CKM phase arises in view of $|E_4^3/E_6^2|\sim 1$.

\smallskip

Furthermore, if the same modular weights are extended to leptons,
$k_L=k_{e_R} = (-6,0,6)$,
all observed lepton masses and mixings can be reproduced with comparable coefficients 
of charged lepton Yukawa couplings and of  effective Majorana neutrino mass operators $(L_iH_u)(L_j H_u)$
such as
\beq
  c^e_{ij} = 10^{-3} \begin{pmatrix}
 0 & 0 & 1.29 \\
 0 & 5.95 & 0.35 \\
 -2.56 & 1.47 & 1.01, 1.32
 \end{pmatrix},
 \qquad
 c^\nu_{ij} = \frac{1}{10^{16}\GeV} \begin{pmatrix}
 0 & 0 & 3.4 \\
 0 & 7.1 & 1.2 \\
 3.4 & 1.2 & 0.19, 0.95
 \end{pmatrix}.
\eeq
CP violation in quarks and neutrinos arises from the unique source $\Re\tau$, but the order
unity unknown factors $c$ prevent precise predictions.
The same structure of $(L_iH_u)(L_j H_u)$ operators
is found applying the modular weights $k_{\nu_R}=(-6,0,6)$ to right-handed neutrinos
and integrating them out, such that $\Re\tau$ can also source baryogenesis via leptogenesis.
With this choice of modular weights the determinant of the 
right-handed neutrino mass matrix also has modular weight 0,
so the resulting  mass matrix of left-handed neutrinos does not contain inverse powers of $E_{4,6}$.
Large mixing angles are here obtained from $c^\nu$ values that compensate
for the mild hierarchy arising from the modular structure.
This could be avoided choosing more equal modular weights for the three-generations of
left-handed lepton doublets $L$.

\smallskip

As an aside comment, we mention that
a simple modular-invariant supergravity potential admits CP-violating minima at $\tau=\pm 0.484+0.884 i$~\cite{2201.02020}.
These values of $\tau$ are near to the special points $\pm e^{\pm 2\pi i/3}$ where $E_4$ vanishes,
so in all our models an order unity CKM phase needs $c^d_{33}/c^{\prime d}_{33}\sim |E_6^2/E_4^3|\approx 2800$, 
and comparable values for all $c^q_{ij}$ coefficients are not possible.

 \subsection{Phenomenology and cosmology}\label{pheno}
The couplings of the modulus $\tau$  (not to be confused with the $\tau$ lepton) to SM particles are predicted. Its components behave similarly to
axion-like particles with special PQ-like charges such that there is no coupling to $G\tilde{G}$.
So the $\tau$ modulus gets no mass from QCD.
Light particles with no QCD anomaly were dubbed `arion' in~\cite{Anselm:1981aw}. 

\smallskip

Moduli were considered problematically light when supersymmetry was expected to exist at the weak scale
for naturalness reasons. Collider data have now shown that this is not the case; 
supersymmetry can exist at much higher energy compatibly with the observed Higgs mass~\cite{1407.4081}.
In such a case, moduli such as $\tau$ can be heavy and decay fast enough to avoid cosmological problems.
The solution to both the hierarchy puzzle and the QCD $\theta$ puzzle can reside in new physics far away from what is currently testable.

Denoting generically as $M_\tau$ the masses of the two $\tau$ scalar eigenstates,
they decay into SM particles such as  $Z q\bar{q}$ conserving baryon number
with width $\Gamma_\tau \sim M_\tau^3/h^2$.
We omitted Yukawa and phase space factors that are presumably dominant for 
$\tau$ decays into heavy right-handed neutrinos.
If the $\tau$ decay constant $h$ is sub-Planckian, $\tau$ can be in thermal equilibrium 
during the big-bang at large temperature, and decouple at 
temperature $T \sim M_\tau$ if $h^2\sim M_\tau \bp$.
So the modulus $\tau$ could have played a role in leptogenesis, but without opening a qualitatively new mechanism for it.
Depending on sparticle masses the fermionic component of $\tau$ might decay slower,
or even be a stable lightest supersymmetric particle, and a Dark Matter candidate.

\smallskip

Having assumed that CP is spontaneously broken only by the  modulus $\tau$, 
its potential is CP-symmetric,
$V(\tau)=V(- \tau^*)$,
and thereby has a pair of degenerate CP-conjugated minima, corresponding to CP broken in opposite directions.
Regions of space in the two minima would be separated by a stable domain wall.
One must assume that CP breaking happened before inflation, so that walls have been inflated away, to avoid
the following problems:
i) the gradient and potential energy of the wall would dominate at late time~\cite{Zeldovich:1974uw,Vilenkin:1984ib};
ii) mechanisms of baryogenesis that rely on this source of CP violation produce opposite-sign asymmetries:
matter on one side and anti-matter on the other side.

\section{Mimicking with  Froggatt-Nielsen models}\label{FN}
We here `deconstruct' the previous model, showing how ordinary Froggatt-Nielsen models based on a U(1)$_{\rm FN}$ symmetry can mimic how
modular symmetries give $\bar \theta=0$ and $\delta_{\rm CKM}\neq 0$.
We denote as $k_\Phi$ the U(1)$_{\rm FN}$ charge of a generic field $\Phi$.
If the U(1)$_{\rm FN}$ symmetry is spontaneously 
broken by a scalar with complex vacuum expectation value $\eta$,
the Yukawa couplings can have the
 forms of eq.\eq{Yij} with modular forms replaced by powers of $\eta$
 \beq F_{k}(\tau) \to \eta^k\eeq 
 and weights replaced by charges.
 Unlike in modular models,
negative powers of $1/\eta$ could be replaced by positive powers of $\eta^*$;
higher order terms would be allowed and contain extra powers of $\eta\eta^*$ which is real.
Despite these differences, $\det Y$ would again be real if the charges of quarks involved in $Y_{ij}$ sum to zero.

However, these Froggatt-Nielsen models lead to vanishing CKM phase:
the Yukawa matrices are only apparently complex, and can be made
explicitly real via appropriate re-phasings of the quark fields, such that $|\eta|$ replaces $\eta$.
Equivalently, one can verify that $\delta_{\rm CKM}\propto \Im\Tr [Y_u^\dagger Y_u, Y_d ^\dagger Y_d]^3$ vanishes.

\medskip

Froggatt-Nielsen models need at least {\em two}
scalar fields with vacuum expectation values with different phases 
to break CP and induce physical CP-violating effects in SM fermions (see e.g.~\cite{0704.0697}).

This feature is built in modular invariance, as
the mathematical properties of modular forms imply that $E_4$ and $E_6$ have different phases.
This property of modular invariance can be mimicked, within Froggatt-Nielsen models,
assuming two scalars $\eta$ and $\eta'$ with charges 4 and 6.
For our purposes there is nothing special in these values.
We can consider Froggatt-Nielsen models with a generic number of scalars $\eta_a$ with charges $k_a$. 
By adjusting quark charges, $\det Y$ can be invariant under U(1)$_{\rm FN}$ phase rotations.
We optimistically assume that the U(1)$_{\rm FN}$ symmetry (if local) is not anomalous; 
that it is negligibly broken  by Planck-suppressed operators;
that charges don't receive quantum corrections. 
The resulting models generate a CKM phase, but extra assumptions are needed to avoid
generating the QCD phase.
Indeed $\det M_q$, while invariant under U(1)$_{\rm FN}$ phase rotations,
can be generically complex in two basic ways:
\begin{itemize}
\item By depending on positive powers of $\eta_a^*$ (such as $\eta \eta^{\prime *}$ if $\eta'$ has the same charge as $\eta$).
These terms could be forbidden by imposing supersymmetry and by building models with potentials such that
super-fields $\eta_a$ are not accompanied by opposite-charge super-fields with non-vanishing vacuum expectation values.

\item By depending on negative powers of $\eta_a$ (such as $\eta /\eta^{\prime}$ if $\eta'$ has the same charge as $\eta$).
Such unwanted terms can arise in QFT models where the fields
$\eta$ also contribute to the masses of mediator particles.
\end{itemize}
Perhaps appropriate U(1)$_{\rm FN}$ models could avoid the above contributions to the QCD angle.
The needed amount of model building highlights the elegance of supersymmetric CP and modular-invariant models, 
where the needed ingredients are built in their mathematical structure:
$E_4(\tau)$ and $E_6(\tau)$ have different phases and terms such as
$E_4^3 E_6^{2*}$ and $E_4^3/E_6^2$ are not included because they are not holomorphic modular forms.
This assumption is critically discussed in the next section~\ref{poles}.

\section{Models with modular functions and anomalies}\label{poles}
So far we have assumed that Yukawa couplings are modular {\em forms}, transforming as
\beq
\label{SLtr}
Y_k(\tau)\to (c\tau+d)^k Y_k(\tau),
\eeq
under SL$(2,\mathbb{Z})$ and holomorphic everywhere in the fundamental domain, including the point $\tau=i\infty$.
Since invariance under the modular group only requires the transformation law in eq.~(\ref{SLtr}) and not
the absence of singularities, we might be tempted to choose as Yukawa couplings modular {\em functions}, that are
singular functions obeying eq.~(\ref{SLtr}).
We now critically analyze the rationale for our assumption.
\smallskip

The only modular form with weight zero is a constant, while the same is not true for modular functions.
The basic modular function with weight 0 is the modular-invariant 
combination of $E_4$ and $E_6$, usually denoted as
\beq
 j(\tau) = \frac{12^3 E_4^3(\tau)}{E_4^3(\tau) - E_6^2(\tau)}\,.
 \eeq
Due to the denominator,  $j$ has a pole at $\tau= i \infty$, so $j$ is a modular {function} but not a modular form.
If Yukawa couplings can depend on  $j(\tau)$ the predictivity of modular flavour models is lost.
Furthermore, $\det M_q$ would be generically complex, 
possibly preventing the understanding of the QCD angle proposed in the present paper.
As we now discuss, models with modular functions can still lead to $\bar\theta=0$.

The choice of discarding modular functions like $j$,
although arbitrary, can be consistently imposed on an effective field theory, since it is mathematically consistent.
Moreover it has a neat physical meaning.
A pole in field space signals that the effective field theory breaks down, 
because some extra state of the full theory with $\tau$-dependent mass become massless at the pole value of $\tau$.
The case of string theory with its infinite number of states will be discussed in section~\ref{string}.
We here focus on  QFT, where by simply including in the theory the extra states that become massless gives
a more complete theory with modular {invariance} realized {through} modular forms.
{Modular invariance} is expected to be exact in the full theory, and can become
apparently anomalous when restricted to light modes in the effective field theory.
This allows to build more general effective field theories with modular functions and QCD anomalies that give rise to $\bar\theta=0$.

As a simple QFT example, right-handed neutrino masses could be modular forms $F_k$ that vanish at specific values of $\tau$.
Integrating out the right-handed neutrinos leads to Majorana $(LH)^2$ neutrino masses operators with a negative power $1/F_k$.
In this example poles indicate points where right-handed neutrinos become massless~\cite{Feruglio:2022kea,Feruglio:2023mii}.
The overall modular weights of $Y^{LL}_{ij}(L_iH_d)(L_j H_d)$ are determined  by those of lepton doublets $L$,
so that large mixing angles can arise assuming $k_{L_1}=k_{L_2}=k_{L_3}$.

\smallskip

Something similar happens in the presence of extra colored heavy fields.
For example a theory with  $3$ generations can be obtained by adding to the MSSM a family $q_H$ and an
anti-family $\bar q_H$ of heavy quarks.
We assume that this is a full theory, where the modular symmetry is non anomalous and realized via modular forms.
The quark modular weights could be $\pm 6, \pm 2, 0$, so that the
sum of the modular weights of all (light and heavy) quarks vanishes.  
The resulting bigger quark mass matrix $M_{\rm all}$
contains a triangular block of 0 entries when written in the basis of increasing  modular weights.
Similarly to the examples in table~\ref{tab:Y000}
this explicitly leads to a real determinant, solving the QCD angle problem.

As an extra step, we discuss how this solution works in the effective theory of light quarks only.
The mass matrix of all quarks can be written as
\beq M_{\rm all} = 
\bordermatrix{ & q_{L \rm light}  & 
q_{L\rm heavy}\hspace{3ex} q^c_{R\rm heavy}  \cr 
q_{R\rm light} & M_{LL} & M_{LH} \cr
\begin{matrix} q_{R\rm heavy} \cr  q^c_{L\rm heavy}  \end{matrix} & M_{HL} & M_{HH}}.
\eeq
The effective theory can be computed in the basis of fixed $q_{\rm light}$ states 
(as they have a non vanishing projection over the light eigenstates), by
integrating out the fixed heavy quarks with $M_{\rm heavy} = M_{HH}$.
The effective mass matrix of the light quarks is
\beq  M_{\rm light} = M_{LL} - M_{LH} M_{HH}^{-1} M_{HL} .\eeq
Its entries have the form of eq.\eq{Yij} with weights dictated by the weights of light fields (kept fixed), 
except that now $F_k$ can be modular {\em functions}
with poles at the specific values of $\tau$ where the heavy quarks become massless.
In practice negative powers of $E_{4,6,\ldots}$ appear if the heavier states are those with higher weight.
The more complicated mass matrices $M_{\rm light}$ of light quarks
facilitate reproducing the observed quark masses and mixings, 
including $\delta_{\rm CKM}\neq 0$.
We see that modular functions describe the effective operators mediated by heavy states with mass comparable to $h$,
the $\tau$ decay constant.
The resulting $\det M_{\rm light}$ is no longer real: because it is a modular function (no longer a modular form),
and because its weight is given by the sum of weights of light quarks only (no longer vanishing).
Matrix algebra shows that it satisfies
\beq \label{eq:detdet}
\det M_{\rm all} = \det M_{\rm light} \det M_{\rm heavy},\eeq
and
\beq
\det M_{\rm light}\to (c\tau+d)^{k_{\rm light}}\det M_{\rm light},\qquad 
\det M_{\rm heavy}\to (c\tau+d)^{k_{\rm heavy}}\det M_{\rm heavy}\eeq
with $k_{\rm light}+k_{\rm heavy}=0$.
In the effective field theory containing the light states only, the
integration over the heavy quarks has produced an additional contribution to the gauge kinetic function $f$:
\beq
f_{\rm IR}=f_{\rm UV}-\frac{1}{8\pi^2}\ln\det M_{\rm heavy},
\eeq
and $\bar\theta=0$ arises as
a cancellation between%
\footnote{We use the notation where 
$f = 1/g_3^2 - i \theta_\mathrm{QCD}/(8\pi^2)$.}
\beq\arg\det M_{\rm light}
\qquad\hbox{and}\qquad
\theta_{\rm QCD}=\arg\det M_{\rm heavy}.\eeq
At the same time, the field content of the low-energy theory is anomalous, but the overall anomaly cancels.
The variation of the path integral measure triggered by a modular transformation 
generates a shift of the gauge kinetic function $f_{\rm IR}$ proportional to $k_{\rm light}$,
exactly compensated by the transformation of $f_{\rm IR}$ under SL$(2,\mathbb{Z})$ :
\beq
f_{\rm IR}\to f_{\rm IR}-\frac{k_{\rm light}}{8\pi^2}\ln(c\tau+d)-\frac{k_{\rm heavy}}{8\pi^2}\ln(c\tau+d)=f_{\rm IR}.
\eeq
Eq.\eq{detdet} neglects higher-dimensional wave-function renormalization factors, 
that are non-negligible if the heavy quarks are not much heavier than the MSSM light quarks.
Again, wave-function factors don't affect $\arg\det M_q$, and the same argument for $\bar\theta=0$ proceeds by replacing
eq.\eq{detdet} with the equivalent exact relation between complex eigenvalues
\beq \prod_i^{\rm all} m_i = \prod_{\ell}^{\rm light} m_\ell \prod_{h}^{\rm heavy} m_h.\eeq
This example shows how the breakdown of the low-energy effective field theory can generate
singularities in the Yukawa couplings, whose origin is related to the appearance of massless modes
in the spectrum of the would-be heavy particles.
In this case, the particle content of the infrared theory is generally anomalous
but the lack of modular invariance due to the light degrees of freedom is 
balanced by the new contribution to the effective theory arising
from the integration of the heavy particles.

\section{Modular invariance in supergravity}\label{sugra}
Planck-suppressed effects could partially spoil the mechanism proposed here,  generating a contribution of order
$\bar\theta \sim h^2/\bp^2$ to the QCD angle.
This would not necessarily be a problem, since we can have $h\circa{<}10^{-5}\bp$:
the $\tau$ modulus is not subject to significant experimental bounds because it does
not need to be light (unlike the axion). In order to analyze the role of such a correction, we extend our study from global supersymmetry to supergravity. 

In supergravity the argument below eq.\eq{Ktransform} gets modified:
the modular variation of the kinetic term for $\tau$ implies a K\"ahler transformation that had no effect in global supersymmetry,
but has an effect in supergravity.
Indeed the supergravity action does not depend on $K$ and $W$ separately, but only on the combination
$G = K / \bp^2+ \ln | W/\bp^3|^2$,
as can be seen via super-Weyl rescalings.
So in supergravity a generic K\"ahler transformation of $K$ needs to be accompanied by a variation of the superpotential
$W$ such that $G$ is invariant:
\beq \label{eq:Ktras} K\to K + \bp^2(F + F^\dagger)\qquad\hbox{and}\qquad W\to e^{-F} W\eeq
where $F$ is a generic dimension-less holomorphic function.
In our case a modular transformation must act on the the superpotential as
\beq W\to (c\tau + d)^{-k_W}W\qquad \hbox{with
modular weight}\qquad k_W = \frac{h^2}{\bp^2} . \eeq
Notice that $k_W$ is necessarily positive. This implies that $W$, evaluated at vanishing values of all matter
multiplets $\Phi$, cannot be a modular form. It has to be a singular modular function.
We come back to this point in the next section.
The new $k_W$ effect is small if $h$ is sub-Planckian, recovering the global supersymmetric limit.\footnote{The models
based on a U(1) Froggatt-Nielsen-like symmetry of section~\ref{FN} can be extended to supergravity without encountering this issue.}
As a result, a cubic coupling $Y$ in the superpotential must transform as
\beq \label{eq:Ymodsug}
Y(\tau)\to  \left(c\tau+d\right)^{k_{q_L}+ k_{q_R}+k_{H_q}-k_W} Y(\tau).\eeq
Furthermore, in supergravity a K\"ahler transformation must be accompanied by a chiral rotation of fermions~\cite{Cremmer:1982en,Witten:1982hu,Derendinger:1991hq}.
This happens because, in the basis where the Einstein term is canonical, the mass terms 
of chiral fermions $\psi^i$ and of gauginos $\lambda$ depend on $K$ as
\beq
-e^{K/2\bp^2} \left[\frac12 ({\cal D}_i D_j W) \psi^i \psi^j - \frac14  K^{i\bar j}
\partial _i f (D_{\bar{j}}  W^\dagger)  \lambda\lambda
\right]\eeq
where $K^{i\bar j}$ is the inverse metric and
\begin{align}
\label{ourL2}
D_i W&=W_i+\frac{K_i}{\bp^2}W\nn\\
{\cal D}_i D_j W&=W_{ij}+\frac{K_{ij}}{\bp^2} W+\frac{K_i}{\bp^2} D_j W+\frac{K_j}{\bp^2} D_i W- \frac{K_i K_j}{\bp^4} W- \Gamma^k_{ij}D_k W\\
\Gamma^i_{kl}&= K^{i\bar h}\partial_k K_{l\bar h}.\nn
\end{align}
Thereby the action is invariant under  the K\"ahler transformation in eq.\eq{Ktras} if it is accompanied by
a U(1)$_R$ phase rotation 
that acts on $\psi$ and $\lambda$ with opposite phase,
preserving the $\lambda\psi \phi^\dagger$ super-gauge interaction:
\beq\label{eq:KahlerChiral}
 \psi \to e^{(F-F^\dagger)/4} \psi,\qquad
\lambda\to e^{-(F-F^\dagger)/4}\lambda.\eeq
(If the K\"ahler space is compact, global consistency of the phase rotation of eq.\eq{KahlerChiral} implies that
$h$ must be quantized in units of $\bp$~\cite{Witten:1982hu}, so $k_W$ is an integer).
Combining these ingredients, a modular transformation acts on canonically normalized matter fields
$\Phi_{\rm can}=\{\phi_{\rm can},\psi_{\rm can}\}$ and gauginos $\lambda$ as a phase rotation
with an extra $k_W$ contribution:
\beq \phi_{\rm can} \to \left(\frac{c\tau+d}{c\tau^\dagger+d}\right)^{-\frac12 k_\Phi} \phi_{\rm can} ,\qquad
\psi_{\rm can}  \to \left(\frac{c\tau+d}{c\tau^\dagger+d}\right)^{\frac14k_W - \frac12k_\Phi}
\psi_{\rm can} , \qquad
\lambda \to \left(\frac{c\tau+d}{c\tau^\dagger+d}\right)^{-\frac14k_W} \lambda.
\label{eq:can}\eeq
The QCD anomaly of the modular transformation is just the sum of the phases in eq.\eq{can},
and acquires a new contribution proportional to $k_W$:
\beq \label{eq:4A}A = \sum_{i=1}^3  \left(2k_{Qi}+ k_{u_{Ri}}+k_{d_{Ri}}  - 2k_W\right)+ C   k_W\eeq
where the term proportional to $C=3$ comes from the gluino.
The product of quark masses relevant for the QCD angle transforms with the same modular weight as the quark anomaly\footnote{This is consistent with the modular transformation of the Yukawa couplings, eq.\eq{Ymodsug}.
A Yukawa coupling in the canonical basis in supergravity is given by
\beq Y_\mathrm{can} = e^{K/2\bp^2} (2\Im\tau)^{\frac12(k_{q_R}+k_{q_L}+k_{H_q})} Y\qquad\hbox{so}\qquad
 Y_{\rm can}\to  \left(\frac{c\tau+d}{c\tau^\dagger+d}\right)^{\frac12 [k_{q_L}+ k_{q_R}+k_{H_q}- k_W \frac{1+ 1}{2}]} Y_{\rm can}.
\eeq}
\beq \det M_q \to \left(\frac{c\tau+d}{c\tau^\dagger+d}\right)^{\sum_{i=1}^3\frac12 (2k_{Q_i}+k_{u_{Ri}}+ k_{d_{Ri}}-2k_W)} \det M_q ,\eeq
having here assumed $k_{H_u}+k_{H_d}=0$, as in section~\ref{CP}.
The gluino mass $M_3$ transforms as
 \beq
 \label{gluinom}
M_3 \to\left(\frac{c\tau+d}{c\tau^\dagger+d}\right)^{\frac12 k_W} M_3 .\eeq
The result  is that the QCD modular anomaly coefficient $A$ is again proportional to the modular weight of the product
of quark and gluino masses relevant for the QCD angle,
\beq \arg \left(M_3^{C}\det M_q\right)\to \left(\frac{c\tau+d}{c\tau^\dagger+d}\right)^{A/2} \arg \left(M_3^{C}\det M_q\right).\eeq
Thus, we might believe that, by choosing weights such that $A=0$, we obtain $\bar\theta=0$ as in the rigid case.
Even assuming that phases in quark and gluino masses are only sourced by modular forms, in supergravity this solution is problematic. The gluino does not mix with quarks and, if the supersymmetry-breaking gluino mass $M_3$ has a positive weight as suggested by eq.~(\ref{gluinom}), this implies that $\det M_q$ must have negative modular weight, leaving some quark massless.

A solution with vanishing anomaly requires a modification of the fermion spectrum.
A minimal extension is adding a chiral color octet multiplet whose fermion $\lambda'$ has modular charge opposite to the gluino, such that the QCD modular anomaly is cancelled,
a modular-invariant real Dirac mass term $\lambda\lambda'$ is allowed,
possibly together with a Majorana mass term $\lambda^2$, so that $\bar\theta=0$.
If $\lambda'$ is heavier than $\lambda$, integrating it out produces an effective theory where
the modular symmetry is anomalous, and $\bar\theta=0$ results from
a cancellation between the gluino complex mass term and the gluon kinetic function, 
analogously to the models of section~\ref{poles}.

\subsection{Solving the QCD $\theta$ problem with anomalous MSSM content}\label{sugramodel}
Going beyond the minimal `Dirac gluino' example we discuss how $\bar\theta=0$
can arise from more general theories that feature more generic extra heavy colored fields, 
such as possibly string theory. 
We here consider full theories where modular invariance is non anomalous,
but a QCD modular anomaly appears in the effective field theory describing the light MSSM fields.
Specifically we choose weights such that $k_{H_u}+k_{H_d}=0$ and 
that the contribution to the QCD modular anomaly of the light quark sector vanishes,
\beq\label{eq:noQanom}
\sum_{i=1}^3  \left(2k_{Qi}+ k_{u_{Ri}}+k_{d_{Ri}}  - 2k_W\right)=0,
\eeq
while the gluino provides an anomaly.
As discussed above, the presence of an anomalous field content is not necessarily a problem, since 
a mechanism for modular anomaly cancellation compatible with a vanishing $\bar\theta$ can be naturally implemented.
In supergravity, gauge anomalies can be cancelled either by an effective one-loop correction to the gauge kinetic function $f$, or by a four-dimensional Green-Schwarz mechanism leading to a one-loop corrected K\"ahler potential $K$~\cite{Derendinger:1991hq}. 
Here we adopt the first option.
We remain with two independent terms in $\bar\theta$, one from the gluino mass term and the other 
one cancelling the anomalous gluino modular transformations. Under reasonable assumptions on the 
supersymmetry breaking sector, the two sum up to $\bar\theta=0$.

\smallskip

As in the example where we have integrated out a heavy sector, our low-energy effective field theory
becomes singular in some region of the fundamental domain. In our case all sources of singularities are related to
the presence of modular functions with negative weight. An example, inherent in the realization of modular invariance
in supergravity, is the superpotential $W$, evaluated at $\Phi=0$. As we have seen, this object is
a modular function with negative weight $-k_W$. 

As discussed in section~\ref{poles} we allow for singularities provided they
have a physical meaning. 
Inspired by typical string theory compactifications, where the limit $\tau\to i\infty$ gives rise to a tower of massless states, we assume that the only singularity in modular functions with negative weight occurs at $\tau= i\infty$. As an extra assumption we also ask the pole at $\tau= i\infty$ to exhibit the mildest possible singularity. 
Under such conditions a modular function  is necessarily proportional to 
$\eta^{2k}$ if $k$ is negative,
$\eta(\tau)= [(E_4^3(\tau)-E_6^2(\tau))/12^3]^{1/24}$ being the Dedekind $\eta$ function.

Finally, we introduce a supersymmetry breaking sector to make gluino massive.
We assume this consists of a new chiral multiplet $S$ invariant under SL$(2,\mathbb{Z})$ such that $\langle D_S W \rangle\ne 0$ with a CP-conserving vacuum expectation value $\langle S\rangle$.
As a minimal realization of this scenario, we consider the K\"ahler potential and the superpotential
\begin{eqnsystem}{sys:KW}  
\label{eq:Kgluino}
K &=&  -k_W\bp^2 \ln(-i\tau+i\tau^\dagger) + \sum_{\Phi} \frac{\Phi^\dagger e^{2 V} \Phi}{(-i\tau + i \tau^\dagger)^{k_\Phi}}-\bp^2\ln(S+ S^\dagger)\\
\label{eq:Kgluinob}
W&=&Y^u_{ij}(\tau) H_{u} u_{Ri} Q_{j} +Y^d_{ij}(\tau) H_{d}  d_{Ri} Q_j+ \frac{c_0 {\bp^3}}{ \eta(\tau)^{2{k_W}}}.
\end{eqnsystem}
Having assumed that the modular transformations of the matter sector are non anomalous,
eq.\eq{noQanom}, we again have real $\det M_q$.
The anomaly related to the
gluino modular transformation is cancelled by requiring that the gauge kinetic function $f$ transforms as
\beq
\label{anof}
f\to f+\frac{k_W}{8\pi^2}C\ln(c\tau+d).
\eeq
To satisfy such a transformation property we can choose\footnote{The Dedekind $\eta$ function transforms with a nontrivial
multiplier system, so the transformations of the matter fields should
also be accompanied by multipliers, as discussed in appendix~\ref{phases}.}
\beq
\label{fkin}
f=f_0S+\frac{k_W}{4\pi^2}  C \ln\eta(\tau).
\eeq
The gluino mass term in supergravity {at tree level is:
\be
\label{gmass}
M_3=\frac{g^2}{2}e^{K/2{\bp^2}} K^{i\bar j}
D_{\bar j} W^\dagger f_i
\ee
and, in our case, can receive contributions by both the dilaton, $(i,\bar j=S,\bar S)$, and the modulus, $(i,\bar j=\tau,\bar \tau)$,  auxiliary fields. However, if supersymmetry is broken dominantly along the modulus direction, squark masses are
non-degenerate and scalar trilinear terms are not aligned with Yukawa couplings, leading to large corrections
to $\bar\theta$, as discussed in section \ref{Corrections}. Thus we are lead to assume that supersymmetry is broken mainly by the dilaton, resulting in tree-level universal squark masses and trilinear terms proportional to Yukawas.
Imposing that $\tau$ preserves supersymmetry gives from eq.s~(\ref{eq:Kgluino},\ref{eq:Kgluinob}) the condition
\be
D_\tau W=- k_W W
\left(
\frac{2 \eta'(\tau)}{\eta(\tau)}+\frac{1}{\tau-\tau^\dagger}
\right)=0.
\ee
Its solutions are $\tau=i$ and $\tau=\pm1/2+i\sqrt{3}/2$, where CP is unbroken and the CKM phase is trivial. 
Nevertheless, $D_\tau W$ can vanish at CP-violating points in the presence of
non-minimal terms in the K\"ahler potential. 
An example is discussed in appendix \ref{Dtau}, where we also compute
the gluino mass for generic  $D_S W$ and $D_\tau W$. 
Assuming spontaneous CP violation and $D_\tau W=0$ at the 
minimum of the scalar potential, eq.~(\ref{gmass}) is enough and gives
\be
\arg M_3=\arg W^\dagger=2k_W \arg\eta(\tau).
\ee
We finally have
\be
\bar\theta=-2 k_W C \arg\eta(\tau)+C \arg M_3=0.
\ee
We arrive at the same result by choosing a basis in field space
where both the contributions from the gauge kinetic function and from the gluino mass separately vanish. This can be achieved by means of the field redefinition:
\beq
\label{fred}
\lambda\to\left[\frac{\eta(\tau)}{\eta(\tau)^\dagger}\right]^{\frac{k_W}{2}}\lambda.
\eeq
Now the chiral transformation of eq.~(\ref{eq:can}) is accounted for by the $\eta$ function and the new gluino field
is modular invariant. In this new basis the gluino mass $M_3$ in eq.~(\ref{gmass}) is real and positive.
At the same time the field redefinition of eq.~(\ref{fred}) is anomalous and generates a new term in the gauge kinetic function:
\beq
\label{anomf}
f\to f-\frac{k_W}{4\pi^2} C \ln \eta(\tau)\equiv f'
\eeq
By combining eq.s~(\ref{anof}) and~(\ref{anomf}) we see that the new gauge kinetic function $f'$ is invariant under
modular transformations (consistently with the new gluino being modular-invariant). 
Assuming it has no singularity, it can be chosen $\tau$-independent:
\beq
f'=f_0 S.
\eeq 
In this new basis, our CP invariant supergravity theory trivially delivers $\bar\theta=0$.

\smallskip

Finally, we can still make use of the phenomenological analysis of section~\ref{concrete}, since
those results can be reproduced by our supergravity theory through a common shift of the quarks modular weights, e.g.\
$k_{\Phi_i}\to k_{\Phi_i}+k_W/2$.

\section{Modular invariance in superstrings}\label{string}
For completeness, we finally recall the string motivation for the mechanism we implemented in QFT, and discuss the possibility of deriving it from string compactification. 
Our solution of the strong CP problem exploits i) a CP-invariant framework, where
CP is spontaneously broken, ii) field-dependent Yukawa couplings shaped by modular invariance
and iii) a possible interplay between ultraviolet and infrared contributions, strongly constrained by anomalies and singularities.
Indeed all these ingredients are naturally present in most string theory compactifications.

\smallskip

First of all, there are strong indications that the four-dimensional CP symmetry is a gauge symmetry
in string theory compactifications~\cite{hep-th/9205011,hep-ph/9205202,hep-ph/9307214}, even starting from a higher-dimensional theory where CP is not conserved. For example, in ten dimensions the heterotic string theory has a charge conjugation symmetry equivalent to an SO(32) (or ${\rm E}_8\otimes{\rm E}_8$) gauge transformation, but has no parity symmetry since the theory is chiral.
In the simplest compactifications, the four-dimensional theory acquires a parity symmetry from a proper Lorentz transformation. Four-dimensional charge conjugation is a combination of a gauge rotation and a proper Lorentz transformation.
Thus, both C and P are gauge symmetries: they arise as combinations of ordinary gauge and general coordinate transformations. Many other compactifications have a gauged CP symmetry. It has been conjectured, as a general property of string theory, that CP is indeed a gauge symmetry of the four-dimensional theory.
In this context, CP can only be violated spontaneously, by complex expectation values of fields.
The problem is to understand why CP violation generated in this way affects dominantly the CKM mixing matrix,
leaving no observable effect in strong interactions. So far, the attempts to solve this problem have mainly focused
either on variants of the Nelson-Barr model~\cite{hep-ph/9307214,1506.05433} or on tuning of the parameters in the low-energy theory~\cite{2002.06931}.

\smallskip

Second, string theory has no free parameters and Yukawa couplings are field-dependent quantities. Their observed value is set by the vacuum expectation values of some scalar fields, the moduli, describing the background over which the string propagates.
Compactifying string theory on suitable spaces with a complex structure, the four-dimensional low-energy effective theory contains generations of chiral fermions with
Yukawa couplings that depend  in a predictive way on such background. Part of this background can be
geometrical and the corresponding moduli describe the shape and the size of the compactified space.
These fields can be seen as Higgs fields that spontaneously break, via compactification, 
flavour and CP symmetries arising from higher-dimensional geometry.
Other moduli include the dilaton and  the extra-dimensional components of the gauge fields.

\smallskip

Third, modular invariance is a key aspect of most string-theory compactifications, allowing a control of
the system beyond perturbation theory. In the energy domain of interest
for present day particle physics, string theory can be approximated by a point particle picture.
The full theory has a spectrum consisting of an infinite number of particles, all 
having a mass of the order of the Planck scale, except a finite number of them.
Integrating out the massive modes leads to an effective theory of the light particles.
This picture resembles that of Kaluza-Klein theories where the low-energy effective action
results from the purely gravitational higher-dimensional system.
Nevertheless, the effective low-energy theory emerging from string theory turns
out to be very special, since it possesses a rich network of discrete duality symmetries \cite{Giveon:1994fu}.

The simplest example of such dualities arises from the compactifications of two extra dimensions on a torus.
In this specific case,
{\em modular invariance} is connected to the way the string perceives the geometry of the compact space:
it can wrap an integer number of times around the cycles of the torus,
and experiences a size $R$ as equivalent to $\sim 1/(\bp^2 R)$.
A torus can be built by dividing the flat complex plane in a lattice 
by identifying $z = z+\omega_1$ and $z=z+\omega_2$ where $\omega_{1,2}$ are two complex constants.
Their absolute values describe the sizes of the two cycles of the torus;
their relative phase describes the twisting angle by which the portion of flat space in one lattice period
is glued at its extremities to form a compact torus. 
This procedure is illustrated in fig.\fig{TwistedTorusConstruction}.
The torus is intrinsically flat, despite it appears curved in the 3d visualisation.
Up to rotations and rescaling (symmetries of string theory), it is described by a modulus $\tau= \omega_2/\omega_1$.
However, this description is redundant,
because $\omega'_2 = a\omega_2 + b\omega_1$, $\omega'_1=c\omega_2 + d\omega_1$
where $a,b,c,d$ are integers such that $ad-bc=1$, give an equivalent lattice.
Similarly to what is done in gauge theories, this redundancy is removed by requiring
the theory to be invariant under the modular group SL(2,~$\mathbb{Z}$).

\begin{figure}[t]
\vspace{-6ex}
$$\includegraphics[width=0.8\textwidth]{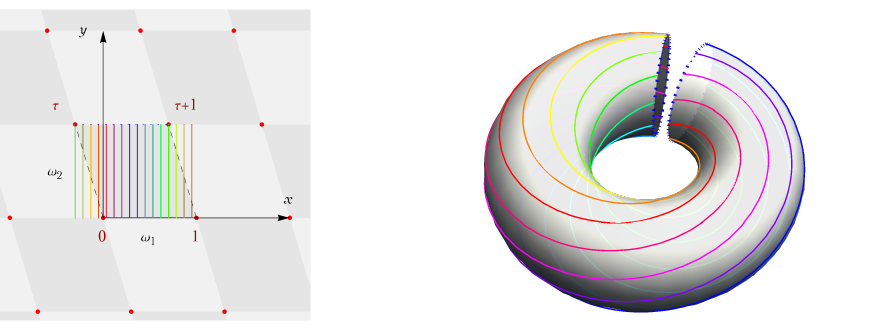}$$
\vspace{-6ex}
\caption{\label{fig:TwistedTorusConstruction} Construction of a flat 2-dimensional twisted torus
by dividing the complex $z=x+iy$ in a lattice $z=z+1$ and $z=z+\tau$ and
identifying opposite points.}
\end{figure}

More generally, the effective action obtained from superstring orbi-folded toroidal compactifications contains multiple moduli 
(moduli $T_i=-i\tau_i$ describe the torus sizes and the fluxes on it;
moduli $U_i$ describe the shapes) with associated symmetries of modular or more complex type.
Phenomenological constructions often focus on a single modulus $\tau$, that can be viewed as
associated to the overall scale of the internal six-dimensional manifold. 
In string compactifications the
modular weights $k$ of fields commonly are integer fractions $k\in \mathbb{Q}$~\cite{hep-th/9202046,Ferrara:1989qb}
and can have both signs, a key feature of our mechanism.

Finally, as in the case of gauge theories, modular invariance should be anomaly-free. However, when integrating out the infinite tower of heavy string states and restricting to the sub-Planckian QFT states,
an anomalous field content can appear in the low-energy theory.
Anomalies are then cancelled by gauge kinetic functions with special dependence on moduli~\cite{Dixon:1990pc,Derendinger:1991hq,hep-th/9202046,Kaplunovsky:1995jw}.
A typical low-energy effective theory associated to an orbifold compactification 
is described by $N=1$ supergravity with K\"ahler potential $K$ and super-potential $W$ given by
\begin{align}
\label{choice}
K=&- \bp^2 \ln(S+S^\dagger)-3 \bp^2 \ln(-i\tau+i\tau^\dagger)+ \sum_i  \frac{\Phi_i^\dagger e^{2V}\Phi_i }{(-i\tau+i\tau^\dagger)^{k_{\Phi_i}}}\nn\\
W=&\frac{\Omega(S) H(\tau) \bp^3}{\eta(\tau)^6}+...
\end{align}
where $S$ is the chiral multiplet containing the dilaton,\footnote{Here we work in a basis where $S$ is modular invariant.} $\tau$ describes here a single overall modulus associated to the volume of the compactified space and $\cdots$ stand for the contribution of matter fields.
The part of the superpotential $W$ displayed above receives nonzero contributions only from non-perturbative string effects~\cite{Ibanez:1990ju,Ferrara:1990ei,Cvetic:1991qm,1812.06520,Leedom:2022zdm}, which
are responsible for the characteristic $\eta(\tau)$ dependence that makes $G=K/\bp^2+\ln|W/\bp^3|^2$ modular invariant.
The function $\Omega(S)$ depends on the specific non-perturbative mechanism, while $H(\tau)$ is a modular-invariant function of $\tau$:
\be
H(\tau)=(j(\tau)-1)^{\beta/2}j(\tau)^{\gamma/3}P_n(j(\tau)),
\ee
where $j(\tau)$ is the Klein absolute invariant, $P_n(j(\tau))$ is a polynomial in $j(\tau)$ of degree $n\ge 0$ and $\beta,\gamma$ are non-negative integers~\cite{rademacher38}. Notice that, since  $j(\tau)$ has a pole at $\tau=i\infty$, the simplest choice $H(\tau)=1$ determines the super-potential with the mildest singularity as $\tau$ approaches $i\infty$.

In general the field content (including weights and possibly phases associated to modular transformations of the fields $\Phi_i$) is anomalous, but the integration over the Planckian modes produces a gauge kinetic function $f$ of the type
\be
f=f_0 S+f_{\rm anomaly}+\cdots
\ee
where $\cdots$ stand for additional modular-invariant contributions, $f_0$ is a constant (here for the sake of illustration we consider a single gauge group $G$) and $f_{\rm anomaly}$ is given by~\cite{Dixon:1990pc,Derendinger:1991hq,hep-th/9202046,Kaplunovsky:1995jw}:
\be
\label{fanomaly}
f_{\rm anomaly}=\frac{1}{4\pi^2}\left[ \sum_\Phi 2 T(\Phi)k_\Phi  + 3\left(C(G)-\sum_\Phi T(\Phi) \right)\right]\ln\eta(\tau).
\ee
The Dedekind $\eta$ function appears summing over string modes
with squared masses $|m+i n T|^2/(T+T^*)$ with integer $n,m$;
each mode gives the usual QFT contribution~\cite{Dixon:1990pc,Ferrara:1991uz,Ibanez:1991qh}.
The term $f_{\rm anomaly}$ cancels the anomaly related to modular transformations~\cite{Kaplunovsky:1995jw,Derendinger:1991hq,hep-th/9202046},
and the expression in brace is exactly $A$ of eq.\eq{4A}, when $G=\SU(3)$ and $k_W=3$.
This is known as 4-dimensional Green-Schwartz mechanism in string theory.
 We see that both non-perturbative effects and anomalies produce a singularity at $\tau=i \infty$ in the effective action. Such a singularity signals the failure of the low-energy effective theory, since in the limit $\tau=i \infty$ 
the theory decompactifies; an infinite tower of states becomes massless and the effective theory is no longer
appropriate to describe the system. Swampland distance conjectures generically suggest poles in Planckian regions of field space~\cite{Ooguri:2006in,1812.06520}.

String constructions thereby contain all the ingredients on which our mechanism to solve the strong CP problem is based. 
On the one side, they appear qualitatively similar to the QFT discussed in section~\ref{poles}, 
where we assumed an anomaly-free modular symmetry thanks to some heavy fields, and
integrated them out. 
On the other side, they can easily embed the model of
section~\ref{sugra}, where both the mixed modular-QCD anomaly and the contribution to $\bar\theta$ 
from the gluino mass are cancelled by the gauge kinetic function.
This leaves hope that the understanding of the QCD $\bar\theta$ puzzle proposed here could be realized
in string constructions.

\section{Corrections to $\bar\theta$}\label{Corrections}
Given the severe experimental bound on $\bar\theta$, special attention should be paid to any source of corrections
potentially affecting the result $\bar\theta=0$, which holds in a specific theoretical limit. 
First, there are well-known corrections to $\bar\theta$ due to the Standard Model dynamics and the
CKM CP-violating phase. 
An additional set of corrections arises from supersymmetry breaking, 
needed to promote our framework into a realistic model. 
Higher dimensional operators, compatible with the symmetry in question, can produce
non-vanishing contributions to $\bar\theta$,
an effect often termed as `quality problem'.

\subsection{Higher dimensional operators}
Regarding the quality problem, the proposed mechanism scores better than the axion~\cite{Luzio} 
and Nelson-Barr solutions to the strong CP problem.

In the limit of unbroken rigid supersymmetry our framework
includes all possible operators depending on the modulus $\tau$, the only source of CP violation. Modular invariance is 
so strong to completely determine the functional dependence of the super-potential on $\tau$, up to a set of real
coupling constants. There is no room for additional modular-invariant operators, provided Yukawa couplings are free from singularities. 

Modular invariance is not so effective to constrain the K\"ahler potential but, as we have seen in section~\ref{CP}, this uncertainty does not impact on $\bar\theta$. 
We have also shown that this conclusion equally applies in the presence of gravitational interactions, at least in the examples we have illustrated in the context of ${ N}=1$ supergravity.

\subsection{Additional sources of CP breaking}
The supergravity models of section~\ref{sugramodel}
include an additional modular-invariant chiral multiplet $S$, sourcing spontaneous supersymmetry breaking. 
To the extent that $S$ has CP-conserving vacuum expectation value, 
no corrections to $\bar\theta=0$ are produced at tree-level. 
For instance, all constants $c^q_{ij}$ in the Yukawa couplings can be
promoted to functions of $S$, without modifying our results.

Moving to string theory, additional gauge singlets of different type are often present in string compactifications.
If they are modular-invariant, we should assume that their vacuum expectation values are CP-conserving. 
This can be easily satisfied, since violating CP through vacuum expectation values of ordinary
modular-invariant scalars requires specific structures with multiple scalars.

\smallskip

In string compactifications, modular invariance under one  SL(2,~$\mathbb{Z}$) group is often extended to
a generalized invariance under multiple $\prod_a$SL(2,~$\mathbb{Z}$)$_a$, with the appearance of 
multiple moduli $\tau_a$.
Our mechanism remains viable if all moduli $\tau_a$
that acquire CP-violating vacuum expectation values
have weights $k^a_{Q,u_R, d_R}$ that satisfy independently in each SL(2,~$\mathbb{Z}$)$_a$ sector
the same condition we assumed e.g.\ in eq.\eq{noQanom}.
One simple possibility is that all moduli but one (our $\tau$) 
acquire CP-conserving vacuum expectation values.
Then no extra conditions are needed.
This scenario does not require a fine tuning, since CP-conserving points of the fundamental domain are good candidates
for extrema of modular-invariant scalar potential~\cite{Font:1990nt,Cvetic:1991qm,Leedom:2022zdm}. 
CP-violating minima exist, but are less easy to obtain~\cite{2201.02020,Leedom:2022zdm}.

\subsection{Standard Model loops}
Corrections to $\bar\theta$ from Standard Model dynamics are known to be negligible, because the SM is invariant under
${\rm U}(3)_Q\otimes {\rm U}(3)_{u_R}\otimes {\rm U}(3)_{d_R}$ redefinitions of quark fields,
and $\bar\theta$ must be a complex invariant under such transformations.
The lowest power of SM Yukawa matrices with the needed properties is 
\beq\label{eq:ThetaSM}
 \bar\theta \propto \Im [(Y_u^\dagger Y_u)^2 (Y_d^\dagger Y_d)^2(Y_u^\dagger Y_u)(Y_d^\dagger Y_d) ]\eeq
times something that differentiates $u$ from $d$ (see e.g.~\cite{hep-ph/9610485}).
Renormalisation-induced effects of this type
arise at 7 loops and contribute as $\bar\theta\sim 10^{-30}$~\cite{Ellis:1978hq}.
The power suppression in eq.\eq{ThetaSM} partially becomes logarithmic when considering IR-enhanced diagrams,
and the largest SM contribution to $\bar\theta\sim10^{-18}$ arises  at 4 loops~\cite{Khriplovich:1985jr}.

\subsection{Supersymmetry breaking}
Larger corrections to $\bar\theta$ can arise at the scale of supersymmetry breaking 
if sparticle masses break CP and/or the ${\rm U}(3)_Q\otimes {\rm U}(3)_{u_R}\otimes {\rm U}(3)_{d_R}$ flavour structure of the SM 
differently from the SM Yukawa couplings.
Whether this happens or not depends on the order between the following mass scales:
\begin{itemize}
\item The scale $\Lambda_{\rm flavour}$ at which the SM or MSSM is replaced by a theory of flavour and CP or,
more in general, by new physics with a different flavour structure, such as SU(5) or SO(10) gauge unification.
In our case $\Lambda_{\rm flavour}$ is the mass scale $M_\tau$ of the modulus $\tau$.

\item The sparticle mass scale, that we generically denote as $m_{\rm SUSY}$.
Based on data and theory we expect that $m_{\rm SUSY}$ is above the weak scale $v=174\GeV$, 
and below the scale at which supersymmetry is broken in some `hidden' sector, that plays no role in the following argument.

\item The mediation scale $\Lambda_{\rm SUSY}$, below which the supersymmetry-breaking soft terms $m_{\rm SUSY}$ 
appear as local operators.  In gauge mediation models~\cite{hep-ph/9801271} $\Lambda_{\rm SUSY}$ is the mass of mediator multiplets.

\end{itemize}
A too large correction to $\bar\theta$ can be avoided if
\beq \label{eq:scales}
m_{\rm SUSY} < \Lambda_{\rm SUSY} < \Lambda_{\rm flavour} .\eeq
In such a limit the set of supersymmetry-breaking  corrections to $\bar\theta$ is not specific to our mechanism based on modular invariance;
rather it is a common property of all supersymmetric  solutions to the strong CP problem relying on spontaneously broken P or CP.
 We can thereby adopt results from previous studies~\cite{Hiller:2002um,Babu:2001se,Hamzaoui:2001uz}. 
The  value
$\bar\theta=0$ set at high-energy $\Lambda_{\rm flavour}$ receives no quantum corrections  down to $\Lambda_{\rm SUSY}$ 
as a consequence of the supersymmetric non-renormalization theorems~\cite{Ellis:1982tk}. 
We are left with corrections below $\Lambda_{\rm SUSY}$, due to RG running of soft terms down to $m_{\rm SUSY}$,
and from integrating out sparticles at the scale $m_{\rm SUSY}$.
Such corrections are model-dependent.

Quark and gluino masses $M_q$ and $ M_3$ receive loop corrections $\delta M_q$ and $\delta M_3$ that induce a correction to
$\bar\theta$
\be
\delta\bar\theta=\delta\bar\theta_q+\delta\bar\theta_g={\rm Im}\Tr(M_q^{-1}\delta M_q)+3\, {\rm Im}\frac{\delta M_3}{M_3}.
\ee
Here we are working in a basis where $\bar\theta=\arg M_3^3 \det M_q=0$.
A correction to the gluino mass $M_3$ can arise from RG effects, while  the threshold correction to $M_3$ decouples  
in the limit $m_{\rm SUSY}\gg v$ of heavy sparticles. 
This is not the case for the threshold correction $\delta\bar\theta_q$, arising from  quark self-energies. 
The leading one-loop result for this quantity is~\cite{Hiller:2002um,Babu:2001se,Hamzaoui:2001uz}:
\beq
\label{Mloop}
\delta\bar\theta_q\sim  \frac{\alpha_3}{4\pi} {\rm Im}\,\tr  \,  \left[ (Y_q)^{-1}
 \sum_{q=u,d} \frac{m^2_{\tilde q_R} }{m_{\rm SUSY}^2}  \left(
\frac{  A_q  }{m_{\rm SUSY}} +  \frac{ v_{q'} }{v_q }Y_q \right) \frac{m^2_{{\tilde q}_L} }{m_{\rm SUSY}^2}
 \right].
\eeq
Here $m^2_{\tilde{q}_{L,R}}$ are the soft mass matrices of left and right-handed squarks $\tilde q$;
$A_{u,d}$ are the trilinear squark/Higgs soft interactions;
$q'=u$ if $q=d$ and viceversa.
So far we have not assumed any particular mechanism of
supersymmetry breaking, and the approximate expression of eq.~(\ref{Mloop}) is generic. 
The correction $\delta \bar\theta_q$ is, in general, dangerously large.

\smallskip

To avoid a too large correction to $\bar\theta$, a commonly invoked assumption is the proportionality between $A_q$ and $Y_q$ together 
with the flavour degeneracy of squark masses.
Corrections from this ideal limit can be computed via a mass-insertion expansion
\beq m_{\tilde{q}}^2 = m_{\tilde{q}}^{(0)2}+\delta m_{\tilde{q}}^2,\qquad
A_q = A_q^{(0)} + \delta A_q,\qquad Y_q = Y_q^{(0)} + \delta Y_q\eeq
where RG effects can be included in the correction terms $\delta$.
Deviations from exact proportionality and/or degeneracy are subject to strong constraints~\cite{Hiller:2002um,Babu:2001se,Hamzaoui:2001uz},
calling for a theoretical justification.
 
The needed structure can be justified assuming that supersymmetry breaking is gauge-mediated~\cite{hep-ph/9801271} 
or anomaly-mediated~\cite{Randall:1998uk,Giudice:1998xp,hep-ph/9912390} at energies below the $\tau$ modulus mass as in eq.\eq{scales}.
In such a case, the RG and threshold corrections due to supersymmetry breaking  have the same flavour and CP structure as the SM corrections, and thereby
undergo the power-like suppression of eq.\eq{ThetaSM}.
This makes the supersymmetric correction to $\bar\theta$ small enough even in the worst case with large $\tan\beta$ and with RG running long enough that
$\ln(\Lambda_{\rm SUSY}/m_{\rm SUSY})$ compensates for the loop suppression $(4\pi)^{-2}$, thereby omitted:
\beq \bar \theta \circa{<}  \frac{M_t^4 M_b^4 M_c^2 M_s^2}{v^{12}} J_{\rm CP} \tan^6\beta \sim 10^{-28} \tan^6\beta . \eeq
 Finally, to avoid a too large $\bar\theta$ at tree level we must assume that the MSSM parameter usually denoted as $B\mu$ is real, otherwise
the Higgses $H_{u,d}$ acquire CP-violating  vacuum expectation values.
Our assumption $k_{H_u}+k_{H_d}=0$ implies a real $\mu$ term.

\section{Conclusions}\label{concl}
The strong CP problem is one of the longstanding puzzles in particle physics.
We addressed it in a plausible theory of CP and flavour motivated by string compactifications:
$N=1$ supersymmetric theories with modular invariance.
We found a neat simple understanding of why $|\bar\theta|\ll 1$ and $\delta_{\rm CKM}\sim 1$,
that also allows to reproduce quark and lepton masses and mixings up to order unity free parameters:
CP broken by the modulus of a non-anomalous modular invariance.
This general scheme has been realized in multiple ways:
\begin{enumerate}
\item In section~\ref{QCDmod} we considered the MSSM with $N=1$ global supersymmetry,
and assumed that the combination of quark modular
weights that controls the QCD modular anomaly sums to zero.
In our simplest example the three generations of quarks have modular weights $-6,0$ and $+6$.
Assuming that the gluino mass is real (for example because supersymmetry is broken by CP-conserving dynamics),
the QCD $\bar\theta$ puzzle is solved as real $\det M_q$ and $\theta_{\rm QCD}=0$, 
assuming that the Yukawa couplings are given by modular forms (modular functions without poles).

\item In section~\ref{poles} we considered
extensions of the MSSM where the modular symmetry is anomaly-free thanks to extra heavy quarks.
For example, adding one vector-like generation, the modular weights could be $\pm 6,\pm 2, 0$.
The QCD $\bar\theta$ puzzle is solved as before.
Furthermore, in the MSSM effective field theory obtained integrating out the heavy quarks, the
modular symmetry is anomalous and Yukawa couplings are given by modular functions,
with poles at the points in field space where the heavy quarks become massless.
In the effective field theory the QCD $\bar\theta$ puzzle is solved as $\bar\theta=\theta_{\rm QCD} + \arg \det M_q=0$.

\item In section~\ref{sugra} we considered supergravity, where the gluino gets unavoidably involved in modular transformations, and contributes to a QCD modular anomaly.
These supergravity effects could either be negligible because Planck-suppressed,
or controlled by dealing with the gluino anomaly similarly to what was done at point 2.
We presented one minimal realization, and one class of models possibly motivated by string compactifications.
\end{enumerate}
In section~\ref{string} we discussed the possibility that the proposed mechanism for $\bar\theta=0$
might be realized in string compactifications, 
and recalled why they provide a plausible motivation for the modular-invariant theories we considered. 
In section~\ref{Corrections} we discussed corrections to $\bar\theta=0$, finding that non-renormalizable operators 
are not problematic, and that (similarly to Nelson-Barr models) supersymmetry breaking must
respect the flavour structure of the SM and be mediated below the flavour scale.
While section~\ref{pheno} discusses phenomenology,
all new particles can be heavy, up to around the Planck scale.

\smallskip

In section~\ref{FN} we discussed if/how the modular understanding for the QCD $\theta$ puzzle
can be realized substituting modular invariance with a spontaneously broken U(1) symmetry {\em a la} Froggatt-Nielsen (FN).
We find that multiple FN scalars are needed to obtain CP-violating Yukawa couplings,
$\delta_{\rm CKM}\sim 1$, and then 
extra ad hoc assumptions are needed to preserve $\bar\theta=0$.
Thanks to its mathematical properties,
modular invariance automatically provides the needed structure,
and behaves like a symmetry automatically broken in a specific way, equivalent to multiple Higgs scalars,
allowing $\delta_{\rm CKM}\sim 1$.
Additionally, FN models need a  mildly small breaking parameter to reproduce quark and lepton masses and mixings up to order one factors.
A mild hierarchy can automatically come from modular invariance, 
as the first non trivial modular forms have weights 4 and 6.

\small

\subsubsection*{Acknowledgments}
This work was supported by the MIUR grant PRIN 2017L5W2PT.
A.S.\ thanks Luis Iba\~nez, Luca di Luzio, Lubos Motl, Paolo Panci and Michele Redi for useful discussions, and 
ChatGPT for proposing the Sanskrit acronym
NAMASTE (Non-Anomalous ModulAr Symmetry for ThEta).

\appendix
\section{Multiplier systems and modular anomalies}\label{phases}
The most general modular transformation of matter fields $\Phi$ reads
\be
\Phi\to e^{-i \alpha_\Phi(\gamma)}(c\tau+d)^{-k_\Phi} \Phi,
\ee	
where $\exp{[-i\alpha_\Phi(\gamma)]}$ is a {\it mutiplier system} depending on the element $\gamma=\{a,b,c,d\}$ of SL(2,~$\mathbb{Z}$). 
In the presence of nontrivial phases $\alpha_\Phi$, the modular invariance of the ${N}=1$ supergravity Lagrangian is still 
guaranteed by the K\"ahler transformation
\be 
\label{eq:Ktras2} K\to K + \bp^2 (F + F^\dagger)\qquad\hbox{and}\qquad W\to e^{-F'} W
\ee
where $F=k_W \ln(c\tau+d)$,
$F'=F+i \alpha_W(\gamma)$ and $\alpha_W(\gamma)$ is an overall phase. This requires a condition on both modular weights and multipliers:
\be\label{cond}
k_{q_{Li}}+k_{q_{Rj}}+k_{H_q}-k_{Y^q_{ij}}=k_W,\qquad
\alpha_{q_{Li}}(\gamma)+\alpha_{q_{Rj}}(\gamma)+\alpha_{H_q}(\gamma)=\alpha_W(\gamma),
\ee
where we took into account that modular forms like $E_{4,6}(\tau)$ have a trivial multiplier equal to one.
The phases $\alpha(\gamma)$ represent a potential source of anomalies, since
a modular transformation acts on canonically normalized fermions
$\psi_{\rm can}$ and gauginos $\lambda$ as a phase rotation with extra, field-independent, contributions:
\beq
\psi_{\rm can}  \to
\left(\frac{e^{-\frac{i}{2} \alpha_\Phi(\gamma)+\frac{i}{4} \alpha_W(\gamma)}}{e^{+\frac{i}{2} \alpha_\Phi(\gamma)-\frac{i}{4} \alpha_W(\gamma)}}
\right)
 \left(\frac{c\tau+d}{c\tau^\dagger+d}\right)^{\frac14k_W - \frac12k_\Phi}
\psi_{\rm can},\qquad
\lambda \to \left(\frac{e^{-\frac{i}{4} \alpha_W(\gamma)}}{e^{+\frac{i}{4} \alpha_W(\gamma)}}
\right)\left(\frac{c\tau+d}{c\tau^\dagger+d}\right)^{-\frac14k_W} \lambda.
\label{eq:can2}
\eeq
Now the QCD anomaly of modular transformations is the sum of two terms,
$A \ln(c\tau+d)+ i A_{\rm phase}$,
where
\begin{align}
A &= \sum_{i=1}^3  \left(2k_{Qi}+ k_{u_{Ri}}+k_{d_{Ri}}  - 2k_W\right)+ C   k_W\nn\\
A_{\rm phase}&=  \sum_{i=1}^3 \left[2\alpha_{Qi}(\gamma)+ \alpha_{u_{Ri}}(\gamma)+\alpha_{d_{Ri}}(\gamma)  - 2\alpha_W(\gamma)\right]+C\alpha_W(\gamma).
\end{align}
Choosing $\alpha_{H_u}+\alpha_{H_d}=0$, eq.~(\ref{cond}) gives $A_{\rm phase}=C\alpha_W(\gamma)$.
If, in line with our mechanism for a real quark determinant, we have
weights satisfying
\be
\sum_{i=1}^3(2 k_{Q_i}+k_{{u_R}_i}+k_{{d_R}_i}- 2k_W)=0,\qquad k_{H_u}+k_{H_d}=0,
\ee
the overall anomaly reduces to $C\left[k_W \ln(c\tau+d)+i \alpha_W(\gamma)\right]$.
The anomaly is cancelled by a gauge kinetic function transforming under SL(2,~$\mathbb{Z}$) as
\be
\label{ftr}
f\to f+\frac{C}{8\pi^2} \left[k_W \ln(c\tau+d)+i\alpha_W\right].
\ee
The modular transformation of the Dedekind $\eta$ function is
$\eta(\tau)\to e^{i\theta(\gamma)}(c\tau+d)^\frac12 \eta(\tau)$ 
with $\theta(S)=- \pi/4$ and $\theta(T)=\pi/12$ for the two generators $S,T$ of the modular group.
So, a choice satisfying eq.~(\ref{ftr}) is
\be
f=\tilde f+ \frac{C}{4\pi^2}k_W\ln\eta(\tau),
\ee
where $\tilde f$ is modular invariant, and the formerly arbitrary phase $\alpha_W(\gamma)$ is fixed to
$\alpha_W(\gamma)=2k_W\theta(\gamma)$. 
Eq.~(\ref{cond}) becomes a constraint on the multiplier systems of matter fields.

\section{The gluino mass}\label{Dtau}
We here compute the gluino mass in the generic case where both $D_S W$ and $D_\tau W$ do not vanish,
explicitly verifying that it has the expected modular transformation properties. 
As the effective supergravity theory contains anomalous terms, that arise at one loop
in the full theory, consistency of the perturbative expansion requires that the
gluino mass is computed at one loop. 
From eq.s~(\ref{sys:KW})
and (\ref{fkin}) we get:
\be
M_3=-\frac{g^2}{2}e^{K/2\bp^2} \frac{W^\dagger}{\bp^2}\left[(S+\bar S)f_0+\frac{C k_W}{8\pi^2}
\left(-i\tau+i\tau^\dagger\right)^2 \left(
\frac{2 \bar\eta'(\tau)}{\bar\eta(\tau)}-\frac{1}{\tau-\tau^\dagger}\right)\frac{2 \eta'(\tau)}{\eta(\tau)}
\right]+\Delta M_3,
\ee
where $\Delta M_3$ is the diagrammatic one-loop contribution~\cite{Bagger:1999rd}:
\be
\Delta M_3= {g^2 \over 16 \pi^2}
\left[\big(3 C - \sum_\Phi T_\Phi\big)m_{3/2}  + 
\big(C - \sum_\Phi T_\Phi\big) \frac{K_S F^S +K_\tau F^\tau}{\bp^2}
+ 2 \sum_\Phi T_\Phi
\left( \ln K_{\Phi\bar\Phi} \right)_\tau F^\tau
\right].
\ee
In this expression, $m_{3/2}=e^{K/2\bp^2} W^\dagger/\bp^2$ is the gravitino mass and $F^i= - e^{K/2\bp^2}\, K^{i \bar j}\, D_{\bar j} W^\dagger$. In our case $\Delta M_3$ evaluates to
\begin{align}
\Delta M_3=
\frac{g^2}{2}e^{K/2\bp^2} \frac{W^\dagger}{\bp^2} \left[ 
\frac{ C}{4 \pi^2} - \frac{C k_W}{8 \pi^2}\,  \left(-i\tau+i\tau^\dagger\right)^2 \left(
\frac{2 \bar\eta'(\tau)}{\bar\eta(\tau)}-\frac{1}{\tau-\tau^\dagger}\right)\frac{1}{\tau-\tau^\dagger}
\right].
\end{align}
Summing the tree and loop terms gives
\be
M_3=-\frac{g^2}{2}e^{K/2\bp^2} \frac{W^\dagger}{\bp^2}\left[(S+\bar S)f_0
- \frac{ C}{4 \pi^2}+\frac{C k_W}{8\pi^2}
(-i\tau+i\tau^\dagger)^2 \left\vert
\frac{2 \eta'(\tau)}{\eta(\tau)}+\frac{1}{\tau-\tau^\dagger}\right\vert^2
\right].
\ee
The required transformation properties of $M_3$ under SL(2,~$\mathbb{Z}$) are correctly reproduced after summing the diagrammatic one-loop contribution to the one coming from the anomaly-modified gauge kinetic function. We also see that the overall phase
of $M_3$ is that of $W^\dagger$.

\smallskip

As discussed in section \ref{Corrections}, we need a mechanism for supersymmetry breaking giving rise to
universal squark masses and trilinear terms proportional to Yukawa couplings. A necessary condition 
is the vanishing of $D_\tau W$ at CP-violating points. To achieve this,
we look for a modification of the K\"ahler potential in eq.~(\ref{eq:Kgluino}), compatible with modular invariance. Focusing on the part that depends only
on the modulus, we consider
\be
K =  -k_W \bp^2 \ln(-i\tau+i\tau^\dagger)+ \alpha(x) \bp^2+...
\ee
where $x$ is the modular-invariant combination $(-i\tau+i\tau^\dagger)|\eta(\tau)|^4$ and $\alpha(x)$ is a real function of $x$ such that the metric $K_{\tau\bar\tau}$ is positive definite. We get
\begin{align}
D_\tau W=&\left[-k_W +x \alpha'(x)\right]\left[
\frac{2 \eta'(\tau)}{\eta(\tau)}+\frac{1}{\tau-\tau^\dagger}
\right]W.
\end{align}
Choices of $\alpha(x)$ exist such that  $-k_W +x \alpha'(x)=0$ and $K_{\tau\bar\tau}>0$ at CP-violating values of $\tau$.
We do not address here the full problem of finding a de Sitter minimum of the scalar potential at such points.

\end{document}